\documentclass[preprint,aps,11pt,preprintnumbers,eqsecnum,nofootinbib]{revtex4}
\usepackage[utf8]{inputenc}
\usepackage{mathtools}
\usepackage{graphics}
\usepackage{graphicx}
\usepackage{dcolumn}
\usepackage{bm}
\usepackage{dsfont} 
\usepackage{amsmath,amssymb}
\usepackage{hyperref}
\usepackage{tabularx}
\usepackage{epstopdf}
\usepackage{tensor}
\usepackage[normalem]{ulem}
\usepackage{footmisc}
\usepackage[usenames]{color}
\hypersetup{
    colorlinks=true,
    linkcolor=blue,
    filecolor=magenta,      
    urlcolor=blue,
    citecolor=blue
}

\newcommand\be{\begin{equation}}
\newcommand\ba{\begin{eqnarray}}
\newcommand\ee{\end{equation}}
\newcommand\ea{\end{eqnarray}}
\newcommand\bw{\begin{widetext}}
\newcommand\ew{\end{widetext}}

\newcommand{\GW}{{\mbox{\tiny GW}}}
\newcommand{\TT}{{\mbox{\tiny TT}}}\newcommand{\T}{\rm T}
\newcommand{\tl}{\rm {t}}
\newcommand{\N}{\rm N}

\newcommand{\LL}{\rm {LL}}
\newcommand{\nn}{\nonumber}

\newcommand{\stkout}[1]{\ifmmode\text{\sout{\textcolor{black}{\ensuremath{#1}}}}\else\sout{#1}\fi}

\begin{document}

\title{Probing Compactified Extra Dimensions with Gravitational Waves}

\author{Yuchen Du}
\affiliation{
 Department of Physics, University of Virginia, Charlottesville, Virginia 22904, USA
}

\author{Shammi Tahura}
\affiliation{
 Department of Physics, University of Virginia, Charlottesville, Virginia 22904, USA
}

\author{Diana Vaman}
\affiliation{
 Department of Physics, University of Virginia, Charlottesville, Virginia 22904, USA
}

\author{Kent Yagi}
\affiliation{
 Department of Physics, University of Virginia, Charlottesville, Virginia 22904, USA
}

\date{\today}

\begin{abstract} 

We study the effect of compact extra dimensions on the gravitational wave luminosity and waveform. We consider a toy model, with a compactified fifth dimension, and matter confined on a brane. We work in the context of five dimensional ($5d$)  general relativity, though we do make connections with the corresponding Kaluza-Klein effective $4d$ theory. We show that the  luminosity of gravitational waves emitted in $5d$ gravity by a binary with the same characteristics (same masses and separation distance) as a $4d$ binary is 20.8\% less relative to the $4d$ case, to leading post-Newtonian order. The phase of the gravitational waveform  differs by 26\% relative to the $4d$ case, to leading post-Newtonian order. Such a correction arises mainly due to the coupling between matter and dilaton field in the effective $4d$ picture and agrees with previous calculations when we set black holes' scalar charges to be those computed from the Kaluza-Klein reduction. The above corrections to the waveform and the luminosity are inconsistent with the gravitational-wave and binary pulsar observations and they thus effectively rule out the possibility of such a simple compactified higher dimensions scenario. We also comment on how our results  change if there are several compactified extra dimensions, and show that the discrepancy with $4d$ general relativity only increases.

\end{abstract}

\maketitle
\bw
\section{Introduction}

Recent discoveries of gravitational waves have opened a new window for testing general relativity (GR), especially in the strong/dynamical field regime~\cite{TheLIGOScientific:2016src,Yunes:2016jcc,TheLIGOScientific:2016pea,Abbott:2018lct,LIGOScientific:2019fpa,Berti:2018cxi,Berti:2018vdi}, in a way which is complementary to other tests including solar system experiments~\cite{TEGP,will-living}, binary pulsars~\cite{stairs,Wex:2014nva} and cosmological observations~\cite{Jain:2010ka,Koyama:2015vza}. Gravitational wave events have been used to probe the fundamental pillars of GR, such as the equivalence principle, Lorentz/parity invariance and massless gravitons~\cite{Yunes:2016jcc}.

One fundamental aspect of gravity that is important to probe is the presence of  extra dimensions motivated by e.g. string theory. For example, in flat $D$-dimensional (non-compact) spacetime, gravitational waves decay with distance travelled as $1/R^{(D-2)/2}$~\cite{Cardoso:2002pa}. This fact has been used to measure $D$ with gravitational wave observations~\cite{Pardo:2018ipy,Abbott:2018lct}. One can also use tidal deformabilities of compact objects to probe extra dimensions. In $4d$ GR tidal deformabilities vanish for non-rotating black holes, but they do not for higher dimensional black holes~\cite{Kol:2011vg,Cardoso:2019vof}.

In this paper we consider gravitational waves in a $5d$ spacetime, with the fifth dimension being compactified on a circle and with the matter constrained on a brane. We work in $5d$ GR rather than in the context of the effective  Einstein-Maxwell-dilaton $4d$ theory (considered in ~\cite{Hirschmann:2017psw,Julie:2017rpw,Julie:2018lfp,Khalil:2018aaj}) which arises from performing a Kaluza-Klein reduction of the $5d$ theory.
Our motivation is two-fold. First motivation is simplicity: it is easier to work with a single field (the metric) than a collection of fields. Also, by not performing the Kaluza-Klein reduction (which assumes that fields are independent of the compactified dimension) allows us to account for the effect of the massive fluctuations that arise when integrating out the compact dimension. By working directly in higher-dimensional GR theory we are also able to generalize the $5d$ results to an arbitrary number of compactified extra dimensions. Second, we hope that our $5$d analysis will be useful to assess the effect of the extra dimensions on gravitational wave detection when considering other paradigms for the geometry of the extra dimensions, e.g. Randall-Sundrum (RS) models.

Gravitational waves in higher dimensional spacetimes with compactified or warped extra dimensions have been studied in the literature. Kaluza-Klein compactification leads to an extra scalar polarization mode(s) (the breathing mode) plus massive Kaluza-Klein modes, whose frequencies are typically much higher than what can be probed with ground-based detectors~\cite{Andriot:2017oaz,Andriot:2019hay}. Such Kaluza-Klein modes also create a stochastic gravitational wave background~\cite{Kwon:2019gsa} and modify the quasinormal modes after mergers~\cite{Toshmatov:2016bsb,Chakraborty:2017qve}. Black hole and neutron star tidal deformabilities have been computed within braneworld models and have been applied to GW170817 to constrain the brane tension~\cite{Chakravarti:2018vlt,Chakravarti:2019aup}. In the RS-II braneworld model~\cite{Randall:1999vf}, black holes may evaporate classically~\cite{emparan-conj,tanaka-conj}, which changes the orbital evolution of binary black holes and further modifies the waveform from that in $4d$ GR. This fact has been applied to the GW events, such as GW150914, to place bounds on the size of the extra dimension~\cite{Yunes:2016jcc}, though such bounds are much weaker than those coming from table-top experiments~\cite{Adelberger:2006dh,Kapner:2006si}. Lastly, given that gravitational waves can propagate in the higher dimensional bulk while electromagnetic waves are constrained on a brane, one can compare the propagation of such two waves to probe the extra dimensions~\cite{Caldwell:2001ja,Abdalla:2001he,Abdalla:2002ir,Abdalla:2002je,Abdalla:2005wr}, which has been applied to GW170817~\cite{Yu:2016tar,Visinelli:2017bny,Yu:2019jlb,Lin_2020}. Again, these bounds on the extra dimension size are much weaker than those from table-top experiments.

 Our paper is organized as follows. In section \ref{setup} we present our conventions, notations and general framework.
In section \ref{sec:Kepler'sLaw}, we extract the compactified $5d$ Newtonian potential and the modified Kepler's law for a binary at a fixed position in the extra dimension.  In section \ref{kkstory} we review the Kaluza-Klein reduction of the $5d$ theory and point out that $5d$ matter, when seen from a $4d$ perspective is non-minimally coupled. Section \ref{metricpert} contains our main result, namely the form of the gravitational waves sourced by the $5d$ binary. 
In section \ref{luminositygw} we extract the luminosity of the gravitational waves, and in section \ref{constraintsgw} we compute the phase of the gravitational waveform as predicted by the compactified $5d$ model, and compare it with observations. Throughout the paper we give generalizations of our formulae in the case of a comapctified $D$-dimensional spacetime, with four non-compact dimensions. We conclude in section \ref{Conclusions}. We relegate some of the more technical details to appendices.

\section{Set-up}\label{setup}

We will consider point-like mass sources in some higher-dimensional spacetime, and we will investigate their effect on the spacetime geometry and on the emission of gravitational waves from binaries, in perturbation theory.

To this end we will compute the metric perturbation $\tilde{h}_{\mu \nu}$ by direct integration of Einstein's equations and not from the  quadrupole formula as it is customary,  because the quadruple formula actually fails when there is a compactified dimension. The reason for this is that the validity of quadrupole formula relies on integration by parts. When the spatial coordinates are non-compact, the boundary terms which accompany the integration by parts are zero. However,  when there are  compactified extra dimensions, there are non-zero boundary terms, which are not straightforward to evaluate and which will contribute, in addition to the usual quadrupole integral. Please see Appendix \ref{appendix: quadrupole formula} for the modified expression.\par
Thus, we need to solve the metric perturbation directly from the Einstein equations. We will  use the  relaxed Einstein equations in the harmonic gauge ~\cite{Pati:2000vt}.

For simplicity in most of this paper we will consider a five-dimensional ($5d$) spacetime with coordinates $x^M$, with four noncompact dimensions $x^\mu$ and a fifth dimension, $x^5=w$, compactified on a circle of radius ${\mathcal R}$, though we will occasionally point out how our results change in the case of additional compactified extra dimensions:
\be
x^M=(x^\mu, x^5,\dots)=(t,\vec x, w,\dots)\sim (t,\vec x, w+2\pi\mathcal R,\dots),\qquad\, M=0,1,2,3,5,\dots D.
\ee
We further denote the spatial coordinates by 
\be
x^I=(\vec x, w,\dots)=(x,y,z,w,\dots), \qquad I=1,2,3,5\dots D\, ,\label{eq:xI}
\ee
and the spatial non-compact coordinates by 
\be
x^i=(x,y,z), \qquad i=1,2,3\,.\label{eq:xi}
\ee
We set the speed of light to $c=1$.

Let us consider a perturbation of the flat spacetime 
\be
h_{MN}\equiv g_{MN}-\eta_{MN}\, , \qquad \eta_{MN}={\rm diag}(-1,1,1,1,1,\dots),
\ee
 and let us also define
\be 
\tilde{h}^{MN}\equiv\eta^{MN}-\tilde g^{MN}\, \qquad \tilde g^{MN}\equiv\sqrt{-g}g^{MN},
\ee
where $(g)$ is the determinant of the metric $g_{MN}.$ As advertised, we take $\tilde h^{MN}$ to satisfy the Lorenz, or de Donder, or harmonic gauge condition:
\be
\partial_M \tilde h^{MN}=0.\label{harmonicg}
\ee
To linear order in $h_{MN}$, $\tilde {h}_{MN}$ reduces to the usual trace-reversed metric perturbation:
\begin{equation}\label{eq:pert2}
\tilde {h}_{MN}\simeq h_{MN}-\frac{1}{2}h\, \eta_{MN}.
\end{equation}
Then, the relaxed Einstein equations state~\cite{Pati:2000vt}
\begin{equation}\label{eq:L0}
\square \tilde {h}^{MN}=-16 \pi G^{(D)} \tau^{MN}\,, 
\end{equation}
where  $G^{(D)}$ is the gravitational constant in the $D$-dimensional spacetime, $\square=\partial_M\partial_N \eta^{MN}$, and
where $\tau^{MN}$ is given by
\begin{equation}
\tau^{MN} = (-g)(T^{MN} + t^{MN}_{\LL} ) +\frac{1}{16\pi G^{(5)}}\bigg({\tilde h^{MP}}{}_{,Q}\,
{\tilde h^{NQ}}{}_{,P}-\tilde h^{PQ} {\tilde h^{MN}}{}_{,PQ}\bigg)\label{tau}
 \,.
\end{equation}
Lastly,  $T^{MN}$ is the matter energy-momentum tensor while $t^{MN}_{\LL}$ is the Landau-Lifshitz ~\cite{Landau:1982dva} gravitational energy-momentum pseudo-tensor. In a $D-$dimensional spacetime we have ~\cite{Yoshino:2009xp} 
\begin{align}\label{eq:landau1}
16 \pi G^{(D)}(-g) t^{MN}_{\LL} &=\tilde{g}\indices{^M^N_{,P}}\tilde{g}\indices{^P^Q_{,Q}}-\tilde{g}\indices{^M^P_{,P}} \tilde{g}\indices{^N^Q _{,Q}}+\frac{1}{2} g^{M N} g_{PQ} \tilde{g}\indices{^P ^R_{,S}} 
\tilde{g}\indices{^Q ^S _{,R}}\nonumber\\
&-\left(g^{M P} g_{Q R} \tilde{g}\indices{^{N R}_{,S}} \tilde{g}\indices{^{Q S}_{, P}}+g^{N P} g_{Q R} 
\tilde g^{MR}{}_{,S} \tilde{g}\indices{^{Q S}_{, P}}\right)
+g_{PQ} g^{R S} \tilde{g}\indices{^{M P}_{, R}} \tilde{g}\indices{^{N Q}_{, S}}\nonumber\\
&+\frac{1}{4(D-2)}\left(2 g^{MP} g^{NQ}-g^{MN} g^{PQ}\right)\left[(D-2) g_{R S} g_{R' S'}-g_{R R' }g_{S S'} \right] \tilde{g}\indices{^{R R'}_{, P}} \tilde{g}\indices{^{S S'}_{, Q}}\,.
\end{align}

From the relaxed Einstein equations \eqref{eq:L0} and the harmonic gauge condition \eqref{harmonicg} it is easy to see that $\tau^{MN}$ obeys  the conservation law
\ba
    \partial_M \tau^{MN}= \partial_M \bigg ((-g) (T^{MN} + t^{MN}_{\LL})\bigg) =0\,.
\ea

\section{Modified Kepler's Law}\label{sec:Kepler'sLaw}

Let us first consider the scenario where there is one extra non-compact spatial dimension. Thus $D=5$, the background is flat, and we assume that there is one matter source which is point-like, of mass $m$, at rest. 
Then, the energy-momentum tensor is
\be
T^{MN}(x^\mu, w)=m \delta^{M0}\delta^{N0}\delta^3(\vec x)\delta(w).
\ee

The only non-trivial linearized metric fluctuation is $\tilde h_{\T}^{00}$, and it satisfies
\be
\square \tilde h_{\T}^{00}=-16 \pi G^{(5)} T^{00}\, ,
\ee
where  $G^{(5)}$ is the gravitational constant in $5d$. The solution is 
\be
\tilde h^{00}_{\T}(\vec x, w)=\frac{4 G^{(5)} m}{\pi (R^2+w^2)}\,, \qquad R^2\equiv \vec x^2 =x^2+y^2+z^2.
\ee
The $5d$ linearized metric fluctuation $h_{\T}^{00}=(2/3)\tilde h_{\T}^{00}$ \footnote{In $D$ spacetime dimensions this relation gets modified to $h_{\T}^{00}=(D-3)/(D-2) \,\tilde h_{\T}^{00}$, and the linearized metric is given by
$ds^2=(-1+(D-3)/(D-2)\,\tilde h_{\T}^{00}) \,dt^2 + (1+1/(D-2) \,\tilde h_{\T}^{00})\,d\vec x\cdot d\vec x.$} corresponds to a Newtonian potential\footnote{Working with a $D$-dimensional spacetime, \eqref{eq:M1} generalizes to
$$
V^{(D)}(R,\rho)=-\frac 12\, \frac{D-3}{D-2}\tilde h_{\T}^{00}=-\frac{D-3}{(D-2)}\frac{4}{\pi^{(D-3)/2}}\Gamma(\frac{D-1}2)\frac{G^{(D)}m}{(R^2+\rho^2)^{D-3}}, \qquad \rho^2=x^I x^I -R^2\, .
$$}
\be\label{eq:M1}
V^{(5)}(R,w)=-{\frac{4}{3}}\frac{G^{(5)}m}{\pi (R^2 +w^2)}\,.
\ee

If the extra dimension is not flat, but compact, with an identification $w\sim w+2\pi \mathcal R$, an observer sees a mass $m$ at every $w=2n\pi \mathcal R$, where $\mathcal R$ is the radius of compactification and $n\in \mathbb{Z}$~\cite{Zwiebach:2004tj}. Summing over all such sources, the resulting  linearized metric fluctuation $\tilde h_{\T}^{00}$ is periodic $\tilde h^{00}_{\T}(\vec x, w) \sim \tilde h^{00}_{\T}(\vec x, w+2\pi n \mathcal R) $:
\be
\label{h00t}
\tilde h^{00}_{\T}(t, \vec x, w)=\frac{4G^{(5)}m}{\pi}\sum_{n=-\infty}^\infty \frac{1}{\vec x^2 + (w-2n\pi \mathcal R)^2}\, ,
\ee
and, correspondingly, the Newtonian gravitational potential is given by
\be\label{eq:M3}
V^{(5,c)}(\vec x, w)=-{\frac{4}{3}}\frac{G^{(5)}m}{\pi}\sum_{n=-\infty}^{\infty}\frac{1}{R^2+(w-2 n \pi \mathcal R)^2}\,.
\ee

 If the observers are located at the same $w$ coordinate as the source (think of the matter source and observer living on the same brane at $w=0$, and ignore for simplicity the backreaction of the brane on the geometry) then we are interested in $V^{(5,c)}(\vec x, w{=}0)$\footnote{For an investigation whether localized matter can arise in the context of effective field theory see \cite{fichet2020braneworld}}. Setting $w=0$ and evaluating the sum over $n$ in \eqref{eq:M3} yields
\be
\label{eq:M22}
V^{(5,c)}(\vec x, w{=}0)=-{\frac{4}{3}}\frac{G^{(5)} m}{l R} \coth\left(\frac{\pi R}{l}\right)=-\frac{1}{3}\tilde h_{\T}^{00}(t, \vec x, w{=}0)\, ,
\ee
where
\be
l = 2\pi \mathcal R\,
\ee
denotes the length of the compactified extra dimension. For a generalization of \eqref{eq:M22} to the case when the observer and the source are located at different positions in the compact dimension, please see Appendix~\ref{hidden}.

There are two useful limits of \eqref{eq:M22}: one is the decompactification limit, when $l \gg R$, and the other is the opposite, with $l \ll R$. In the first case, the Newtonian potential assumes the form of a $4d$ non-compact \emph{space} 
\be
V^{(5,c)}(R,w{=}0) = -{\frac{4}{3}}\frac{G^{(5)} m}{\pi R^2}+{\cal O}(R/l)\, , \qquad R=\vec x^2=x^ix^i\, , \qquad l\gg R\,,
\ee
whereas in the second case it is equal to the Newtonian potential in a $3d$ non-compact space plus exponential corrections\footnote{
For  a $D$-dimensional space time $R^{3,1}\times T^{D-4}$, with three non-compact spatial dimensions and the compact space being torus, the generalization of \eqref{eq:Ma2} is 
$$V^{(D)}(x^I) = -\frac{ 2(D-3)}{D-2} \frac{G^{(D)} m} {{\rm Vol(Compact\;Space)} } \frac 1R \bigg(1+ {\cal O}(e^{-2\pi R/l}) \bigg)\, , 
$$
where $l$ is the length of the largest of the cycles of the torus. }
\be
V^{(5,c)}(R,w{=}0) = -{\frac{4}{3}}\frac{G^{(5)} m} {lR} \bigg(1+2e^{-2\pi R/l} + {\cal O}(e^{-4\pi R/l}) \bigg)\,, \qquad l\ll R .\label{eq:Ma2}
\ee
The exponential corrections look like a Yukawa potential, and can be interpreted as being due to massive gravitons.
From a $4d$ perspective, these massive gravitons correspond to non-uniform Fourier modes of the massless $5d$ gravitons on the circle $w\sim w+l$. We will have more to say about this in the following sections.

Next, let us consider a quasi-circular binary with component masses $m_1$ and $m_2$ and binary separation $r_{12}$, with $r_{12}\gg l$. The matter energy-momentum tensor is given by
\be
T^{MN}(x)=\sum_{a=1,2} m_a \int d\tau \frac{\dot x_a^M \dot x_a^N}{\sqrt{-g_{PQ}(x) \dot x_a^P \dot x_a^Q}} \delta^5(x - x_a(\tau))\,,\label{bin0}
\ee
where $x_a^M(\tau)$ is the trajectory of the point-like mass $m_a$ with $\tau$ representing an affine parameter on the worldline. We can use reparametrization invariance to identify $x^0(\tau)=\tau$ and, assuming that the matter sources are located at $w_1(t)=w_2(t)=0$ (i.e. confined to the same brane), to leading order we have 
\be
\vec x_{12}(t) = \vec x_1(t) -\vec x_2(t)= (r_{12} \cos(\Omega t), r_{12}\sin(\Omega t), 0)\,.\label{bin1}
\ee
Further using \eqref{eq:Ma2} yields the effective potential of such a binary 
\be
V_{\text{eff}}\simeq \frac{1}{2} \mu r_{12}^{2} \Omega^{2}-\frac{G_{\N}\mu M}{r_{12}}\left(1+2 e^{-2\pi r_{12} / l}\right)\,,\label{bin2}
\ee
 where
\be
M=m_1+m_2\label{total}
\ee
 is the total mass of the binary and 
\be
\mu=(m_1m_2)/M\label{reduced}
\ee
 is the reduced mass, while $\Omega$ is the orbital angular frequency. $G_{\N}$ is the $4d$ Newton's constant, with\footnote{ For  a $D$-dimensional space time $R^{3,1}\times T^{D-4}$, with three non-compact spatial dimensions and the compact space being torus, the $4d$ Newton's constant is given by
$$G_{\N} =\frac{ 2(D-3)}{D-2} \frac{G^{(D)} }{\rm Vol(Compact\;Space)},$$
with $G^{(D)}$ the $D$-dimensional gravitational constant.
}
\be G_{\N}\equiv {\frac{4}{3}}\frac{G^{(5)}}{l}\, .\label{gnewton}
\ee 
 The distance between the two sources is solved from the condition of local extremum of $V_{\text{eff}}$ with respect to $r_{12}$. This leads to the following modification to the Kepler's law:
\ba
r_{12} \simeq \left(\frac{G_{\N} M}{\Omega^{2}}\right)^{1 / 3}\left\{1+\frac{2}{3}\left(\frac{G_{\N} M}{\Omega^{2}}\right)^{1/3} \frac{2\pi }{l}\exp\left[-\frac{2\pi }{l}\left(\frac{G_{\N} M}{\Omega^{2}}\right)^{1/3}\right]+\frac{2}{3}\exp\left[-\frac{2\pi }{l}\left(\frac{G_{\N} M}{\Omega^{2}}\right)^{1/3}\right]\right\}\, ,\nn\\
\ea
where we have retained the first order correction to the $4d$ Kepler's law.

\section{Performing the Kaluza-Klein reduction with $5d$ point-like matter sources}\label{kkstory}

One might be tempted to think that the physics of the binary system in a $5d$ spacetime is that of a binary (two point-like masses) coupled to the fields obtained via Kaluza-Klein reduction of the $5d$ metric, namely gravity, dilaton and Maxwell fields, and with the latter two being set to zero. Then, to leading order, neglecting all corrections coming from massive modes on the fifth dimensional circle, one recovers the $4d$ matter (the binary) plus gravity set-up.
However, this is not the case. To better understand this issue, we take a quick detour and review the  Kaluza-Klein reduction of the $5d$ system composed of gravity plus point-like sources. This is a self-contained section of the paper and for the purpose of performing the Kaluza-Klein reduction we introduce the notation $G_{MN}$ for the $5d$ metric and $g_{\mu\nu}$ for the $4d$ metric.

Consider five-dimensional gravity 
\be
{\cal S} _{\rm 5d}= \frac{1}{l \kappa} \int d^5x \sqrt{-{\rm{det}}G_{MN}} \,R[G_{MN}],
\ee
 where $\kappa=16\pi G$ (we work in units where $c=1$) and 
$
G=G^{(5)}/l.
$
  The Kaluza-Klein reduction ansatz to $4d$ is (see for example \cite{Duff:1986hr}):
\be
G_{MN} = e^{-\varphi/3}
\begin{pmatrix}
g_{\mu\nu}+\kappa e^{\varphi}A_\mu A_\nu&\sqrt{\kappa} e^{\varphi}A_\mu\\
\sqrt{\kappa} e^{\varphi} A_\nu&e^{\varphi}
\end{pmatrix}, \label{kk}
\ee
where all the fields in (\ref{kk}) are functions of the 4d coordinates, $x^\mu$, only.

Substituting (\ref{kk}) into the 5d Einstein-Hilbert action yields the 4d Einstein-Maxwell-dilaton action
\be
{\cal S}_{\rm KK} =\frac{1}{\kappa} \int d^4 x \sqrt{-g} \bigg(R[g_{\mu\nu}] -\frac \kappa 4 e^{ \varphi} F_{\mu\nu} F^{\mu\nu} -
\frac{1}{6} \partial_\mu \varphi \partial^\mu \varphi\bigg)\,.\label{kk red action}
\ee

Note that we can find solutions with a vanishing dilaton as long as the Maxwell field is pure gauge. (The dilaton equation is sourced by the Maxwell field, so setting the dilaton to zero in general would lead to an inconsistent Kaluza-Klein truncation.)

By introducing the rescaled dilaton and Maxwell field as
\be
\label{eq:rescale}
\varphi = - 2\sqrt{3} \phi\,, \quad A_\mu = \frac{2}{\sqrt{\kappa}} \bar A_\mu\,,
\ee
we can rewrite the Einstein-Maxwell-dilaton action as
\be
{\cal S}_{\rm KK} =\frac{1}{\kappa} \int d^4 x \sqrt{-g} \bigg(R[g_{\mu\nu}] -e^{-2\sqrt{3} \phi} \bar F_{\mu\nu} \bar F^{\mu\nu} -
2 \partial_\mu \phi \partial^\mu \phi\bigg)\,.
\ee
This agrees with~(II.1) of~\cite{Julie:2017rpw}. We will use the rescaled  dilaton \eqref{eq:rescale} to make contact with~\cite{Julie:2017rpw} in Section~\ref{luminositygw}.

Consider now adding a point-like source of mass $m$ to the $5d$ action:
\be
{\cal S}_{\rm matter,\,5d}=-m\int d\tau \sqrt{-\dot x^M(\tau)\dot x^N(\tau) G_{MN}(x(\tau))}\, ,\label{matter}
\ee
where $\tau$ is an affine parameter on the source's worldline and $\dot x^M=\frac{d}{d\tau}x^M$.
This will source the $5d$ metric in the usual way, leading to the $5d$ perturbative analysis performed in the previous section, and continued in the next.

Here we would like to point out that the $4d$ dilaton is also being sourced by the $5d$ matter (\ref{matter}).
Specifically, for a source that is not moving in the fifth dimension (note that this is a solution to its equation of motion in the context of a $5d$ metric which is independent of the fifth coordinate), the reduction of the $5d$ action (\ref{matter}) yields
\be
{\cal S}_{\rm matter,\, KK}{=}-m  \int d\tau \,e^{- \varphi/6}\sqrt{-\dot x^\mu(\tau) \dot x^\nu(\tau) (g_{\mu\nu}+\kappa e^{ \varphi} A_\mu A_\nu)} \,, \label {matter kk}
\ee
where the $4d$ fields are evaluated on the worldline.

In contrast, adding a $4d$ neutral source of mass $m$ to the Einstein-Maxwell-dilaton action is done by considering
\be
{\cal S}_{\rm {matter},\,4d}=-m \int d\tau\sqrt{-\dot x^\mu(\tau) \dot x^\nu(\tau) g_{\mu\nu}}\, .
\ee
The main message here is that $5d$ matter couples not only to the $4d$ graviton, but also to the dilaton and Maxwell field. Most importantly, while we can find solutions with a vanishing dilaton to the $4d$ action with a $4d$ matter source, we cannot find solutions with a vanishing dilaton to the $4d$ Kaluza-Klein reduced action when the matter source is a $5d$ point-like source; this can be understood by noticing that in (\ref{matter kk}) there is a linear coupling between the dilaton and the matter source,  when expanding in small fluctuations about a vanishing dilaton. In effect, the  $4d$ mass in the Kaluza-Klein reduced action is modulated by the dilaton, 
\be
m_{\rm effective}= m \exp(-\varphi/6)\,, \label{eq:m_eff}
\ee as evidenced by (\ref{matter kk}).

Lastly, in  the context of Kaluza-Klein reduction of a higher-dimensional gravitational theory, the $4d$ gravitational coupling constant is $1/(16 \pi G)$, as we can see from \eqref{kk red action}, with \be
G=G^{(5)}/l\,.\label{ggrav}
\ee
However, $G$  and $G_{\N}$ (which shows up in the Newtonian potential and it is given in \eqref{gnewton} for $D=5$) are not equal: they differ by a factor. This is different from $4d$ GR, where $G$ and $G_{\N}$ are equal to one another. The explanation for this mismatch stems from the fact that in an effective $4d$ theory, the interaction between two masses is not only gravitational, but there are additional contributions mediated by $4d$ scalars (e.g. dilaton) as well.

\section{Metric perturbations}\label{metricpert}
We now return to our main problem, namely finding the $5d$ metric fluctuations sourced by a binary in a spacetime with one compact dimension of radius $\mathcal R$.

We separately find the contribution of the matter energy-momentum tensor and of the Landau-Lifshitz pseudo-tensor to the metric fluctuations: we denote these by $\tilde h^{MN}_{\T}$ and $\tilde h^{MN}_{\tl}$. There is one more contribution to the metric fluctuations from the remainder of the relaxed Einstein equation source $\tau^{MN}$ (see \eqref{tau}). However, to leading order, the extra terms in $\tau^{MN}$ contribute only to the 00-component of the metric fluctuations, $\tilde h^{00}$. We will compute the 0-components of the metric fluctuations not by direct integration, but by using the harmonic gauge \eqref{harmonicg}. So, for the remaining fluctuations $\tilde h^{IJ}=(\tilde h^{ij}, \tilde h^{i5},\tilde h^{55})$ we will evaluate first the contribution from the matter source, and then use this in the Landau-Lifshitz pseudo-tensor to evaluate the non-linear metric fluctuations. Despite $\tilde h^{IJ}_{\tl}$ being non-linear, it is actually of the same order as $\tilde h^{IJ}_{\T}$ in a velocity expansion. Lastly, we add the two contributions to find the metric perturbation to second order in velocities, i.e. leading order in post-Newtonian expansion.

\subsection{Metric perturbations: contribution from the matter sources}\label{mpT}

We begin by computing the perturbations sourced by the matter stress-energy tensor. The energy-momentum tensor $T^{MN}$ of a system of point masses at $w=0$ is given by
\begin{align}\label{eq:L1}
T^{MN}(t,\vec x, w)=\sum_{a} 
\frac{P^{M}_{a} P^{N}_{a}}{P_{a}^0} \delta^{3}\left(\vec x-
\vec x_a(t)\right)\delta\left(w\right)\,,
\end{align}
where $P^{M}_a=m \dot x_a^M /\sqrt{-g_{PQ}  \dot x_a^P \dot x_a^{Q} }$ is the $M$-component of the momentum of particle $a$. We parametrize the particles' trajectories with $\vec x_a=\vec x_a(t) $ and take $x_a^0=t$. For a binary source, we specifically use (\ref{bin1}).

From \eqref{eq:L0}, the linearized fluctuations are given by 
\begin{align}\label{eq:L2}
\tilde{h}^{MN}_{\T}(t, \vec x,w)=-16 \pi G^{(5)} \int dt^{\prime}d^3\vec x{}^{\prime}dw^{\prime}\, \mathcal G^{(5,c)}\left(t,\vec x,w; t', \vec x^{\prime},w^{\prime}\right) T^{MN}\left(t^{\prime},\vec{x}{}^{\prime}, w^{\prime}\right)\,,
\end{align}
where $\mathcal G^{(5,c)}(t,\vec x,w;t', \vec x^{\prime},w^{\prime})$ is the (scalar) retarded compactified Green's function in $5d$.
The retarded Green's function in flat $5d$ can be represented as~\cite{Dai:2013cwa}
\begin{align}\label{eq:L3}
\mathcal G^{(5)}\left(t, \vec x,w; t',  \vec x^{\prime},w'\right)=-\frac{ \theta(t-t^{\prime})}{4 \pi^2 r}\frac{\partial}{\partial r} \frac{\theta\left(t-t^{\prime}-r\right)}{\sqrt{\left(t-t^{\prime}\right)^{2}-r^{2}}}\,,
\end{align}
with 
\be
r^2=(\vec x-\vec x')^2+ (w - w^{\prime})^2\, ,
\ee 
and where $\theta$ denotes the Heaviside step-function. \footnote{For the reader accustomed to $4d$ expressions, we want to point out that even though the $5d$ retarded Green's function does not have support only on the light-cone (as  opposed to the massless $4d$ retarded Green's function which has support on the light-cone only), it does have support  inside the light-cone, and it is therefore causal. This is one of the peculiar features of odd dimension spacetimes. }

Then, starting from \eqref{eq:L3}, we can write the compactified $5d$ retarded Green's function, $\mathcal G^{(5,c)}(x,y)$:
\begin{align}\label{eq:L3c}
\mathcal G^{(5,c)}\left(t, \vec x,w; t',  \vec x^{\prime},w'\right)=-\sum_{n=-\infty}^\infty \frac{\theta(t-t')}{4\pi^2 r_n} \frac{\partial}{\partial r_n} \frac{\theta\left(t-t^{\prime}-r_n\right)}{\sqrt{\left(t-t^{\prime}\right)^{2}-r_n^{2}}}\,,
\end{align}
where $r_n^2=(\vec x-\vec x')^2+(w-w'-nl)^2$.
For practical purposes, the compactified Green's function expression given in \eqref{eq:L3c} is not very useful.\footnote{However, for the purpose of demonstrating how one could use \eqref{eq:L3c} in an explicit calculation, please see Appendix~\ref{NpotintretG} for another derivation of the Newtonian potential in $5d$ GR.}
Instead, we will use the equivalent representation of  the compactified retarded Green's function in terms of a sum/integral over Fourier modes\footnote{The Dirac-delta function, written as a distribution on the space of  periodic functions with period $l$, is $\delta(w-w')=(1/l) \sum_{n=-\infty}^\infty \exp( i 2 \pi n  (w-w')/l) $.} (see also Appendix \ref{Green}): 
\be
 \mathcal G^{(5,c)}(x^\mu, w;x'{}^\mu, w')=-\frac 1l \sum_{n=-\infty}^\infty
\int \frac{d^4 p}{(2\pi)^4} \frac{e^{i p\cdot (x-x')} e^{i 2\pi n (w-w')/l} }
{-(p_0+i\epsilon)^2+\vec p^2+(2\pi n/l)^2}\, , \label{fourier exp ret G}
\ee
where $\epsilon$ is an infinitesimally small positive number.
In \eqref{fourier exp ret G} each term in the sum can be interpreted as the $4d$ retarded Green's function of a massive particle, of mass $m_s= (2\pi n/l)$. These massive particles are nothing else but the massive Kaluza-Klein graviton states. Thus we expect that in the limit when $r\gg l$, and for slow moving sources, \eqref{fourier exp ret G} will reduce to the $4d$ retarded Green's function of a massless particle, corresponding to $n=0$, plus exponentially suppressed corrections, with the leading order correction coming from the least massive mode, corresponding to $n=1$. Indeed, the $n=0$ term in the sum above corresponds to massless $4d$ excitations, and the retarded $4d$ Green's function
$\theta(t-t') \delta(t-t'-|\vec x-\vec x'|)/(4\pi |\vec x-\vec x'|)$. The non-zero $n$ terms are associated with massive $4d$ excitations. The retarded Green's function for a massive $4d$ scalar of mass $m_s$  is 
\begin{align}
-\int \frac{d^4 p}{(2\pi)^4} \frac{e^{i p\cdot (x-x')}  }
{-(p_0+i\epsilon)^2+\vec p^2+m_s^2}&=-\frac{\theta(t-t')}{4\pi}\bigg[
 \frac{\delta (t-t' -|\vec x- \vec x'|)}{|\vec x - \vec x'|} \nonumber\\&-
 \theta (t -t'-|\vec x-\vec x'|) \frac{m_s J_1 \left( m_s  \sqrt{(t-t') ^2- |\vec x-\vec x'|^2}
   \right)}{\sqrt{(t-t')^2 - |\vec x-\vec x'|^2}} \bigg]\,.
\end{align}
Consider next  the propagation of a periodic signal $e^{i \omega t'}f(\vec x')$, with $f(\vec x')$ localized near the origin (similar to the case encountered with the binary sources). In the leading multipole expansion, for $
|\vec x-\vec x'|\simeq |\vec x|=R$ we are left with evaluating
\begin{align}
& \int^t_{- \infty} {d}t'\ \left\{ \frac{\delta (t - t' - R)}{R} -
   \theta (t - t' - R) \frac{m_s J_1 \left( m_s  \sqrt{(t - t')^2 - R^2}
   \right)}{\sqrt{(t - t')^2 - R^2}} \right\} e^{i \omega t'} \nn\\
  &=  \frac{e^{i \omega (t - R)}}{R} - m_s e^{i \omega t} I_{\frac{1}{2}}
  \left[ \frac{R}{2} \left( \sqrt{m_s^2 - \omega^2} - i \omega \right) \right]
  K_{\frac{1}{2}} \left[ \frac{R}{2} \left( \sqrt{m_s^2 - \omega^2} + i \omega
  \right) \right] \nonumber \\
 & 
  = \frac{e^{i \omega t}}{R} e^{- R \sqrt{m_s^2 - \omega^2}}\label{eqn:master}\, .
\end{align}
If $m_s \gg \omega$ (which is the case for slow moving binary sources
since $m_s = \frac{2 \pi n}{l}, l \ll
r_{12}, \Omega r_{12} \ll 1$), the approximate result from \eqref{eqn:master} would be simply $(1/R)\,{e^{i \omega
t}} e^{- 2 \pi n \frac{R}{l}}$, which is the anticipated exponentially suppressed contribution.

So, putting everything together, the signal propagating from a source that is localized near the origin $f(\vec x', w') e^{i \omega t'}$ to a spacetime coordinate $(t,\vec x, w)$ is
\begin{align}
&\int dt' \int d^3 \vec x' \int_0^l dw'\,\mathcal G^{(5,c)}(t,\vec x, w; t',\vec x', w') f(\vec x',w')e^{i\omega t'}\simeq
-\frac{e^{i\omega (t-R)}}{4\pi l R}  \int d^3 \vec x'\int_0^l dw'\, f(\vec x', w') \nn\\
&\qquad-\sum_{n, n\neq 0}
\frac{e^{i \omega t}}{4\pi l R} e^{- R \sqrt{(2\pi n/l)^2 - \omega^2}}\int d^3 \vec x'\int_0^l dw'\,f(\vec x', w')e^{2\pi i n (w-w')/l}\, .\label{signal}
\end{align}
In the limit of a small extra-dimension and a slow moving source (i.e. $2\pi/l\gg \omega$), we find that the leading contribution is 
\be
-\frac{1}{4\pi l R}  \int d^3 \vec x'\int_0^l dw'\, f(\vec x', w') e^{i\omega (t-R)}\,,\label{leading}
\ee
which corresponds to a signal that propagates uniformly in $w$ and radially in the non-compact space. The massive Kaluza-Klein gravitons give an exponentially suppressed contribution of the form
\be
-2\sum_{n=1}^\infty
\frac{e^{i \omega t}}{4\pi l R} e^{- 2\pi n R/l}
\int d^3 \vec x'\int_0^l dw'\, f(\vec x', w') \cos(2\pi n (w-w')/l) \,.\label{subleading}
\ee
At large distances $R\gg l$ these massive states contributions can be safely ignored.

In particular, from \eqref{eq:L2}, sourced by the binary (\ref{bin0}, \ref{bin1}) energy-momentum, and using the approximations in \eqref{leading} and \eqref{subleading} in the far-field slow motion limit, we find for example the $(x,y)$ component as
\begin{align}\label{hxyT}
 \tilde{h}^{xy}_{\T} (t,\vec x, w{=}0) 
    \simeq &-{\frac{3}{2}}\frac{G_{\N} \mu}{R} r_{12}^2 \Omega^2 \left(\sin[2\Omega(t-R)] + \sum_{n=1}^{\infty} e^{-\frac{2\pi R}{l} n} \sin(2 \Omega t)\right)\,,
\end{align}
where $R=|\vec x|$ is the $3d$ distance between the sources and the observer, and $\mu$ is the reduced mass defined in \eqref{reduced}.
The leading correction due to the extra compact dimension to the part of the metric fluctuation which is sourced by the matter energy-momentum tensor is given by the $n=1$ term in the sum in \eqref{hxyT}, and it  is an exponentially suppressed correction. Since $l\ll R$, the correction $\exp({-\frac{2\pi R}{l}})$ is extremely small and can be safely ignored (after all we have already ignored corrections of order $r_{12}/R$, and we expect $l< r_{12}$).

\subsection{Metric perturbations: the non-linear contribution from the Landau-Lifshitz pseudo-tensor}\label{mptl}

We denote the non-linear metric perturbations sourced by $t_{\LL}^{MN}$ in \eqref{eq:L0} as  $\tilde{h}_{\tl}^{MN}$. In the slow motion limit ($v\ll 1$), since $\tilde h^{00}_{\T}\sim{\cal O}(1)$, $\tilde h^{0i}_{\T}\sim{\cal O}(v),$  $\tilde h^{ij}_{\T}\simeq {\cal O}(v^2)$ and $\tilde h^{M 5}_{\T}\simeq 0$, the leading order contribution for $I,J=1,2,3,5,\dots D $ comes from 
\begin{align}\label{eq:nonlinear}
    \tilde{h}_{\tl}^{IJ}(x) &= -16\pi G^{(D)}\int d^5y\, \mathcal G^{(D,c)}(x,y) t_{\LL}^{IJ}(y) \nonumber \\ 
    &\simeq - \frac{D-3}{4(D-2)} \int d^5y\, \mathcal G^{(D,c)}(x,y) 
 \partial_{M}\tilde{h}^{00}_{\T}(y)\partial_N \tilde{h}^{00}_{\T}(y) (2 \eta^{I M} \eta^{JN} - \eta^{IJ}\eta^{ MN}),\end{align}
where $\mathcal G^{(D,c)}(x,y)$ is the compactified retarded Green's function in $D$ dimensions.  Specializing to the case $D=5$, we get
\begin{align}\label{eq:nonlinear}
    \tilde{h}_{\tl}^{IJ}(x) 
    &\simeq - \frac{1}{6}  (2 \eta^{I M} \eta^{JN} - \eta^{IJ}\eta^{MN})\int d^5y\, \mathcal 
G^{(5,c)}(x,y) 
 \partial_{M}\tilde {h}_{\T}^{00}(y)\partial_N \tilde{h}^{00}_{\T}(y),\end{align}
where $\mathcal G^{(5,c)}(x,y)$ was  previously defined in \eqref{eq:L3c} and \eqref{fourier exp ret G}. As discussed in the previous subsection, in the far field limit (when the distance to the source is much larger than the distances between sources) with the observer and the sources  located at $w=0$, and in the slow motion approximation,
the compactified retarded $5d$ Green's function reduces effectively to a $4d$ retarded Green's function. The first order correction, which is proportional to  $\exp({-\frac{2\pi R}{l}})$ is negligible, and \eqref{eq:nonlinear} becomes
\be
\label{eq:nonlinear1}
 \tilde{h}_{\tl}^{IJ}(t,\vec x,0) \simeq  \frac{  1}{4\pi l R}\frac{1}{6}  (2 \eta^{IN} \eta^{JM} - \eta^{IJ}\eta^{MN})\int d^3  y \int _0^l dw \, \partial_{M}\tilde {h}^{00}_{\T}(t-R,\vec y,w)\,\partial_N \tilde{h}^{00}_{\T}(t-R,\vec y, w). 
\ee
The most striking difference in the non-linear contribution to the metric fluctuations in $5d$ with respect to the $4d$ case  is the coefficient $1/6$ on the right hand side of \eqref{eq:nonlinear1} relative to the more familiar coefficient of $1/8$ in $4d$. 

We now return to the specific case of a binary system at $w=0$, with masses $m_1,m_2$ moving in the $(x^1,x^2)$ plane. The first observation is that in \eqref{eq:nonlinear1}, to leading order in velocities we only need to to consider the action of the spatial derivatives on $\tilde h_{\T}^{00}$, which does not contain an explicit $t$-dependence. Restricting now the sumation over $M,N$ indices to spatial indices $K,L$, consider the term $\partial_K \tilde{h}_{\T}^{00} \partial_L \tilde{h}^{00}$ in~\eqref{eq:nonlinear1}. Since $\tilde{h}_{\T}^{00} = \tilde{h}_{\T,1}^{00} + \tilde{h}_{\T,2}^{00}$, there will be four terms. However, we are only interested in the two crossing terms because non-crossing terms will be simply regularized and effectively be dropped out. In addition, we will replace the spatial derivative on $y$ to the derivative with respect to the position of the sources (with a minus sign). We can do so because of translation invariance of the flat background which implies that the linearized fluctuation $\tilde{h}^{00}_{\T,a}$ only depends on $\vec y-\vec y_a$ and $w-w_a$. We will use $\partial_K^{(a)}$ to represent partial derivatives with respect to the coordinates of the source $a$, ${\partial}/{\partial y_a^K}$. With the help of this little trick, we can simplify \eqref{eq:nonlinear1}:
\begin{align}\label{eq:nonlinear2}
    \tilde {h}^{IJ}_{\tl}(t,\vec x,w) &\simeq\frac{(2 \eta^{IK} \eta^{JL} - \eta^{IJ}\eta^{KL})(\partial_K^{(1)}\partial_L^{(2)} + \partial_K^{(2)}\partial_L^{(1)})}{24 \pi  l R} \int_{\rm{NZ}}d^3 y \int_{0}^{l} dw\  \tilde h_{\T,1}^{00}(t-R,\vec y, w) \tilde h_{\T,2}^{00}(t-R,\vec y, w)\, ,
\end{align}
where $\int _{\rm{NZ}} d^3 y$ denotes integration in the near zone (NZ) region (i.e. in the vicinity of the sources) which is the region that contributes the most to the volume integral  $\int d^3 y$~\cite{Pati:2000vt}. Due to the near-zone approximation, the wave propagation is almost instantaneous and we can use for $\tilde h^{00}_{\T,a}$  in the NZ region the result from \eqref{h00t},
\begin{align}\label{eq:h00-2}
    \tilde{h}_{{\rm{NZ}},\T,a}^{00} (t, \vec y,w)= -\frac{4}{\pi} G^{(5)} m_a \sum_n \frac{1}{(\vec y-\vec x_a(t))^2 + (w - w_a+n l)^2}\,.
\end{align}
Substituting \eqref{eq:h00-2} into the integrand \eqref{eq:nonlinear2}, we can use the infinite sums to extend the integration region  over $w$ first, and then we can perform the remaining sum exactly:
\begin{align}\label{eq:nonlinear3}
 \int_0^l dw'\, \tilde{h}_{{\rm{NZ}},\T,1}^{00} (t, \vec y,w')    \tilde{h}_{{\rm{NZ}},\T,2}^{00} (t, \vec y,w') &= \sum_{n_1, n_2}\int_0^{l} dw'\  \frac{16  m_1 m_2 (G^{(5)})^2}{\pi^2(R_1^2 + (w' - w_1 + n_1 l)^2)(R_2^2 + (w' - w_2 + n_2 l)^2)}\nonumber\\
    &=  \int_{-\infty}^{+\infty}d w' \sum_{n_2} \frac{16m_1 m_2(G^{(5)})^2}{\pi^2(R_1^2 + w'^2)(R_2^2 + (w' + w_1 - w_2 + n_2 l)^2}\nonumber\\
    &= \frac{16 m_1 m_2 (G^{(5)})^2(R_1 + R_2)}{\pi R_1 R_2}\sum_{n_2}  \frac{1}{(R_1 + R_2)^2 + (w_1 - w_2 + n_2 l)^2}\nonumber\\
    &= l \frac{9 m_1 m_2 G_{\N}^2 }{R_1 R_2} \frac{\sinh \frac{2\pi(R_1+R_2)}{l}}{\cosh\frac{2\pi(R_1 + R_2)}{l} - \cos\frac{2\pi
(w_1 - w_2)}{l}}\nonumber\\
    &\simeq l \frac{9 m_1 m_2G_{\N}^2 }{R_1 R_2} \left(1 + 2 e^{-\frac{2\pi(R_1 + R_2)}{l}} \cos\frac{2\pi(w_1 - w_2)}{l}\right)\, ,
\end{align}
where $R_1=|\vec y-\vec x_1(t)|$,  $R_2=|\vec y - \vec x_2(t)|$, and where $G_{\N}$ was previously defined in \eqref{gnewton}. In the last step in \eqref {eq:nonlinear3} we used $l \ll r_{12 }\leq R_1 + R_2$, with $r_{12}$ the binary separation distance.
It is important to keep the explicit $w_i$ dependence because we will still have to take derivatives respect to the position of the sources in the $5d$ spacetime. Substituting \eqref{eq:nonlinear3} into \eqref{eq:nonlinear2} we obtain
\begin{align}
\label{eq: h from t}
     \tilde{h}^{IJ}_{\tl}(x) 
     &\simeq \frac{ 3}{ 2} \frac{m_1 m_2G_{\N}^2 }{2 \pi R} (2 \eta^{IK} \eta^{JL} - \eta^{IJ}\eta^{KL})(\partial_K^{(1)}\partial_L^{(2)} + \partial_K^{(2)}\partial_L^{(1)})\left(- \pi r_{12}+ l \, e^{-\frac{2\pi r_{12}}{l}} \cos\frac{2\pi(w_1 - w_2)}{l}\right)\,.
\end{align}
For more details on how the integration was performed in \eqref{eq: h from t}, please see  Appendix \ref{App:DirectvsQuad}.

In particular, from \eqref{eq: h from t}, for a binary at $w_1=w_2=0$ as in \eqref{bin0} and \eqref{bin1}, we obtain e.g. the $(x,y)$ component of the metric perturbation as
\begin{align}
    \tilde {h}^{x y}_{\tl}  (t,\vec x, 0)\simeq 
    & -\frac{3}{ 2 } \frac{ m_1 m_2 G_{\N}^2 }{R} (\partial_1^{(1)}\partial_2^{(2)} + \partial_1^{(2)}\partial_2^{(1)}) \left(r_{12} -\frac l\pi e^{-\frac{2\pi r_{12}}{l}}\right) \nonumber\\
    \simeq & -\frac{ 3}{ 2}\frac{m_1 m_2 G_{\N}^2 }{R r_{12}}\left(1 + 2e^{-\frac{2\pi r_{12}}{l}} + \frac{4\pi r_{12}}{l}e^{-\frac{2\pi r_{12}}{l}}\right) \sin[2 \Omega (t-R)]\,.
\end{align}
We can further use the modified Kepler's law $\Omega^2 = \frac{G_{\N} M}{r_{12}^3}(1 + 2e^{-\frac{2\pi r_{12}}{l}} + \frac{4\pi r_{12}}{l}e^{-\frac{2\pi r_{12}}{l}})$ to cast it into a more familiar form:
\be
\tilde{h}^{x y}_{\tl} (t,\vec x, 0)\simeq -\frac{3}{2} \frac{G_{\N} \mu}{R} r_{12}^2 \Omega^2 \sin[2\Omega(t-R)]\,.
\ee

\subsection{Gravitational Waves from a Binary Source in a 5d spacetime}

Similar calculations to the ones we presented in explicit detail in the previous sections yield the following expressions for the other non-zero linearized fluctuations $\tilde{h}_{\T}^{IJ}$  (sourced by the matter energy-momentum tensor), as well as the leading order non-linear fluctuations, $\tilde h_{\tl}^{IJ}$ (sourced by the Landau-Lifshitz pseudo-tensor):
\begin{align}\label{bunchofhs}
    \tilde{h}^{xx}_{\T} (t, \vec x, 0)\simeq {3}\frac{G_{\N}\mu}{R}r_{12}^2 \Omega^2 \sin^2[\Omega(t-R)], \qquad&\tilde{h}^{xx}_{\tl}(t, \vec x, 0)\simeq -{ 3}\frac{G_{\N} \mu}{R} r_{12}^2 \Omega^2 \cos^2[\Omega(t-R)]\,,\nn\\
    \tilde{h}^{yy}_{\T} (t, \vec x, 0)\simeq {{3}}\frac{G_{\N}\mu}{R}r_{12}^2 \Omega^2 \cos^2[\Omega(t-R)], \qquad &\tilde{h}^{yy}_{\tl}(t, \vec x, 0) \simeq -{ 3}\frac{G_{\N} \mu}{R} r_{12}^2 \Omega^2  \sin^2[\Omega(t-R)] \,,\nn\\
   \tilde{h}^{zz}_{\T}(t, \vec x, 0) \simeq 0, \qquad &\tilde{h}^{zz}_{\tl}(t, \vec x, 0)\simeq 0\,,\nn
\\
\tilde h_{\T}^{ww}(t, \vec x, 0) \simeq 0, \qquad &\tilde{h}^{ww}_{\tl}(t, \vec x, 0)\simeq -{3}\frac{ G_{\N} \mu}{R}r_{12}^2 \Omega^2 \left(1 - 4\frac{2\pi r_{12}}{l}e^{-\frac{2\pi r_{12}}{l}} \right)\, ,
\end{align}
where we recall that $\mu$ is the reduced mass of the binary \eqref{reduced}.
(As a caveat, we would like to point out  that we cannot set $z_1 = z_2 = w_1 = w_2 = 0$ until the derivatives in \eqref{eq: h from t} have been taken, and the vanishing of $\tilde{h}_{\tl}^{zz}$ is not trivial.) As advertised, both $\tilde{h}_{\T}^{IJ}$ and $\tilde h_{\tl}^{IJ}$ are of the same order in velocities.

To second order in velocities, the non-zero metric fluctuations $\tilde h^{IJ}$  are obtained by adding the linearized $\tilde h_{\T}$  and non-linear $\tilde h_{\tl} $:
\begin{align}\label{eq:binarypert1}
&\tilde{h}^{x y}(t, \vec x, 0) =\tilde{h}^{x y}_{\T} + \tilde{h}^{x y}_{\tl} \simeq - {3}\frac{G_{\N} \mu}{R} r_{12}^2 \Omega^2 \sin[2\Omega(t-R)]\,,\nn\\
    &\tilde{h}^{xx} (t, \vec x, 0) = \tilde{h}^{x x}_{\T} + \tilde{h}^{x x}_t \simeq -{3}\frac{G_{\N} \mu}{R}r_{12}^2 \Omega^2 \cos[2\Omega(t-R)]\,,\nn\\
    &\tilde{h}^{yy} (t, \vec x, 0) = \tilde{h}^{yy}_{\T} + \tilde{h}^{yy}_{\tl} \simeq  {3}\frac{G_{\N} \mu}{R}r_{12}^2 \Omega^2 \cos[2\Omega(t-R)] \,,\nn\\
&\tilde h^{ww}(t, \vec x, 0)= \tilde{h}^{ww}_{\T} + \tilde{h}^{ww}_{\tl} \simeq -{3}\frac{ G_{\N} \mu}{R}r_{12}^2 \Omega^2 \left(1 - 4\frac{2\pi r_{12}}{l}e^{-\frac{2\pi r_{12}}{l}}\right)\, .
\end{align}
According to our discussion in Sections \ref{mpT} (see \eqref{subleading}) and \ref{mptl}  (see \eqref{eq: h from t}), the metric fluctuations  
$ \tilde h^{IJ}(x^\mu, w)$ are equal to $\tilde h^{IJ}(x^\mu,w{=}0)$, up to exponentially suppressed corrections.
However, the biggest change to the luminosity of the gravitational waves and the phase of the gravitational waveform comes from the leading order terms  retained in \eqref{eq:binarypert1}. 

The extension of the results given in \eqref{eq:binarypert1} to a compactified $D$-dimensional spacetime is straightforward: 
\begin{align}
\tilde{h}^{(D)}{}^{I J} = \frac{D - 2}{2 (D - 3)}
\left(\begin{array}{cc}
          \tilde{h}^{(4)}{}^{i j} & 0\\
        0 &\Big(-4\frac{G_N \mu}{R} r_{12}^2 \Omega^2\Big)
        \delta^{p q}
      \end{array}\right), \qquad i, j = 1,2,3 \;\;{\rm and} \;\; p,q =5,6,\dots D \,,
\end{align}
{where $\tilde{h}^{(4)}{}^{ij}$ denotes the $4d$ gravitational waves sourced by a binary with the same characteristics as ours: reduced mass $\mu$, separation distance $r_{12}$, angular frequency $\Omega$, and located at $z=0$:
\begin{align*}
\tilde{h}^{(4)}{}^{i j} = \left(\begin{array}{ccc}
        -{4}\frac{G_{\N} \mu}{R}r_{12}^2 \Omega^2 \cos[2\Omega(t-R)]& -{4}\frac{G_{\N} \mu}{R} r_{12}^2 \Omega^2 \sin[2\Omega(t-R)]& 0\\
        -{4}\frac{G_{\N} \mu}{R} r_{12}^2 \Omega^2 \sin[2\Omega(t-R)]& 4\frac{G_{\N} \mu}{R}r_{12}^2 \Omega^2 \cos[2\Omega(t-R)]& 0\\
        0& 0& 0
      \end{array}\right), \qquad i, j = 1,2,3 \,.
\end{align*}}

Lastly, the remaining metric fluctuations $\tilde h^{0M}$ can be obtained either by direct integration, or more easily, by using the harmonic gauge \eqref{harmonicg} condition.
In the next sections we will use the latter.

\section{The Luminosity of Gravitational Waves}\label{luminositygw}

In this section we compute the luminosity of gravitational waves. We work in the harmonic gauge \eqref{harmonicg}, without specializing to the  more commonly used transverse-traceless gauge (for a comparison, see Appendix \ref{app:TT}). 

There is one more subtlety we would like to comment on before we begin. In the $D$-dimensional gravitational theory, the only coupling constant is the gravitational constant $G^{(D)}$ of the Einstein-Hilbert action.
After performing the Kaluza-Klein reduction, the effective $4d$ theory has the gravitational constant $G=G^{(D)}/{\rm Vol(Compact\;Space})$ and  the Newton's constant is $G_{\N}=(2(D-3)/(D-2))G$. In contrast, in a strictly $4d$ theory of gravity coupled to matter, we would have $G=G_{\N}$. 

In our subsequent comparisons between the predictions of the compactified higher-dimensional gravity theory and $4d$ GR we will identify the Newton's constants in the two theories. 

We use the gravitational energy-momentum pseudo-tensor $t_{\LL}^{MN}$ given in~\eqref{eq:landau1}. Since $t_{\LL}^{MN}$ is already second order in the metric fluctuations, we can use the linearized  approximation for $\tilde h^{MN}$, ~\eqref{eq:pert2}, to obtain
\begin{align}\label{eq:landau2}
16 \pi G^{(D)}t^{MN}_{\LL}&\simeq\tilde{h}\indices{^M^N_{,P}}\tilde {h}\indices{^P^Q_{,Q}}-\tilde{h}\indices{^M^P_{,P}}\tilde{h}\indices{^N^Q_{,Q}}+\frac{1}{2}\eta^{MN}\tilde{h}\indices{^P^R_{,Q}}\tilde{h}\indices{^Q_{P,R}}-\left(\tilde{h}\indices{^M^P_{,Q}}\tilde{h}\indices{_P^Q^{,N}}+\tilde{h}\indices{^N^P_{,Q}}\tilde{h}\indices{_P^Q^{,M}}\right)\nn\\&
+\tilde{h}\indices{^M^P^{,Q}}\tilde{h}\indices{^N_P_{,Q}}+\frac{1}{2}\tilde{h}^{PQ,M}\tilde{h}\indices{_P_Q^{,N}}-\frac{1}{4}\eta^{MN}\tilde{h}^{PQ,R}\tilde{h}_{PQ,R}\nn\\&-\frac{1}{4(D-2)}\left(2\tilde{h}^{,M}\tilde{h}^{,N}-
\eta^{MN}\tilde{h}^{,P}\tilde{h}_{,P}\right)\, , 
\end{align}
where the indices are raised and lowered with the Minkowski metric.
Further imposing the Lorenz gauge \eqref{harmonicg}  and performing short-wavelength averaging\footnote{When performing short-wavelength averaging, integration by parts is permitted~\cite{Brill:1964zz}.}, ~\eqref{eq:landau2} becomes
\be\label{eq:landau3}
\langle( t_{\LL})_{MN} \rangle \simeq\frac{1}{32 \pi G^{(D)}}\left\langle \partial_{M}\tilde{h}_{PQ}\partial_{N}\tilde{h}^{PQ}-\frac{1}{D-2}\partial_{M}\tilde{h}\partial_{N}\tilde{h}\right\rangle\,.
\ee
The total energy carried away by gravitational waves is given by the following volume integral:
\begin{equation}
E_{\GW}=\int d^3\vec x\int_0^l dw \, t_{\LL}^{00}(t,x^I)\simeq l\int d^3\vec x \,t_{\LL}^{00}(t,\vec x)\, ,
\end{equation}
where in the last step we used that, to leading order, the metric fluctuations propagate uniformly in $w$.
Then the rate of change of the radiated energy is
\begin{align}\label{eq:Edot}
\dot{E}_{\GW}=\frac{dE_{\GW}}{dt}=&\int d^3\vec x\int_0^l dw \,\partial_0 t_{\LL}^{00}\nonumber\\
=&-\int  d^3\vec x\int_0^l dw \, \partial_I t_{\LL}^{I0}\nonumber\\
=&\oint dA \int_0^l dw  \,(t_{\LL})_{0I}n^{I}\,,
\end{align}
where we recall that our index conventions defined in \eqref{eq:xI} and \eqref{eq:xi} are: $I,J=1,2,3,5$  and $i,j = 1,2,3$.
In \eqref{eq:Edot}, $dA$ is the differential area element on the  2-sphere  at spatial infinity and $n^I$ is the unit vector along the direction of propagation of the gravitational waves. From \eqref{leading} and \eqref{eq:binarypert1} we saw that the gravitational waves propagate radially in the non-compact directions and uniformly in $w$ to leading order. The non-uniform propagation along the direction of compactification is due to the massive Kaluza-Klein modes which yield exponentially suppressed corrections. So, to leading order, the only non-zero components of $n^I$ are $n^i$. Then, the rate of change of energy in a 2-sphere at a distance $R$ from the source becomes
\begin{equation}\label{eq:Edot2}
\dot{E}_{\GW}=l \oint d\Omega\,(t_{\LL})_{0k}n^k R^2 \,.
\end{equation}
Using repeatedly  the harmonic gauge and the fact that the perturbations in the far zone depend on the retarded time, we obtain the following identities to leading order in $1/R$:
\ba\label{eq:hbar-id}
&\partial_k\tilde{h}_{IJ} \simeq -\dot{\tilde{h}}_{IJ} n_k\,, \nn \\
&\partial_k\tilde{h}_{00}\simeq -\dot{\tilde{h}}_{ij}n^in^jn_k\,,\nn \\
&\partial_k \tilde{h}_{0I}\simeq \dot{\tilde{h}}_{Ij} n^jn_k\,,\nn \\
&\dot{\tilde{h}}_{00}\simeq \dot{\tilde{h}}_{ij} n^in^j\,,\nn\\
&\dot{\tilde{h}}_{0I}\simeq -\dot{\tilde{h}}_{Ij} n^j\,,
\ea
where a dot denotes a time derivative. In more detail, in writing $\partial_k\tilde h_{IJ} \simeq -\dot{\tilde{h}}_{IJ} n_k$, we used the fact that the metric fluctuations are spherical waves (see \eqref{eq:binarypert1}), and to leading order in $1/R$, we can ignore the action of  $\partial_k$ derivative on the $1/R$ factor. Then, when acting on the periodic function of $t-R$, we can trade off $\partial_k$ for $n_k \partial_R$ and the latter for $-n_k \partial_t$.

Substituting  \eqref{eq:hbar-id} into~\eqref{eq:landau3} we derive the follwing result 
\ba\label{eq:landau5}
\langle t_{0k}n^k \rangle \simeq -&\dfrac{1}{32\pi G^{(D)}}\left\langle\dot{\tilde h}_{IJ}\dot{\tilde  h}^{IJ}+\frac{D-3}{D-2}\dot{\tilde h}_{ij}\dot{\tilde h}_{kl}n^in^jn^kn^l\right. \left.-2\dot {\tilde h}_{Ij}\dot {\tilde  h}^{Ik}n^jn_k-\frac{1}{D-2}\dot {\tilde h}\indices{^I_I}\dot{\tilde h}\indices{^J_J}\right.\nonumber\\ & \left.+\frac{2}{D-2}\dot{\tilde h}\indices{^I_I}\dot{\tilde h}_{ij}n^in^j\right\rangle \,.
\ea

We are now ready to compute the luminosity of the gravitational waves from a binary source.  For $D=5$, substituting the perturbations derived in \eqref{eq:binarypert1} into \eqref{eq:Edot2}, and noting that to leading order we have $\partial_0  \tilde h^I{}_I=\partial_0\tilde h^i{}_i=0$, we find \footnote{{
For general $D$-dimensions,
\begin{align*}
\dot{E}_{\GW}^{(D,c)}
\simeq -\frac{7D-16}{15(D-2)}\frac{R^2}{8G_N}\frac{2(D-3)}{D-2}\left\langle \dot {\tilde h}_{ij}\dot {\tilde h}^{ij}\right\rangle 
&= -\frac{7D-16}{15(D-2)}\frac{D-2}{2(D-3)} 16 G_N^{7/3} \mu^2 M^{4/3} \Omega^{10/3}\\
&= \frac{7D-16}{12(D-3)}\dot{E}_\GW^{(4)}\,,
\end{align*}
where $\dot{E}_\GW^{(4)}$ is defined in \eqref{eq:luminosity4d}}\label{EdotD}.}
\ba \label{eq:Edot3}
\dot{E}_{\GW}^{(5,c)}\simeq -\frac{19}{360}\frac{R^2}{G}\left\langle \dot {\tilde h}_{ij}\dot {\tilde h}^{ij}\right\rangle\,,
\ea
where we recall that $G$ is the effective $4d$ theory gravitational constant \eqref{ggrav}, and we used the isotropy of the gravitational waves together with the following identities:
\ba 
\int d^2\Omega\, n^i n^j &=&\frac{4\pi}{3}\delta^{ij}\,,\nn\\
\int d^2\Omega\, n^i n^j n^k n^l &=&\frac{4\pi}{15}\left(\delta^{ij}\delta^{kl}+\delta^{il}\delta^{jk}+\delta^{ik}\delta^{jl}\right)\,.
\ea
Substituting the metric perturbations derived earlier in \eqref{eq:binarypert1} into \eqref{eq:Edot3}, and keeping terms only to leading order in velocity, the compactified $5d$ GR luminosity is
\be \label{eq:luminosity}
 \dot{E}_\GW^{(5,c)}\simeq -\frac{304}{{45}}G \mu^2 r_{12}^4\Omega^6={-\frac{76}{15} G_{\N}^{7/3}} \mu^2 M^{4/3} \Omega ^{10/3}\,.
\ee
In contrast, the luminosity of gravitational waves in a purely $4d$ gravitational theory, with the gravitational waves sourced by a binary with the same characteristics as ours, is equal to
\ba \label{eq:luminosity4d}
\dot{E}_\GW^{(4)}\simeq -\frac{32}{5}{G_{\N}} \mu^2 r_{12}^4\Omega^6=-\frac{32}{5} {G_{\N}}^{7/3} \mu ^2 M^{4/3} \Omega ^{10/3}\,.
\ea
We conclude that the  luminosity of gravitational waves in a $5d$ spacetime with a compact fifth dimension differs by ${20.8\%}$ from the corresponding $4d$ GR luminosity.

Let us now compare the luminosity derived earlier in \eqref{eq:luminosity} with the predictions of Einstein-Maxwell-dilaton theory studied in Refs.~\cite{Julie:2017rpw,Julie:2018lfp,Khalil:2018aaj}. For neutral matter (i.e.  the electric charges are zero), the energy of a binary is dissipated  via gravitational and scalar (dilaton)  radiation. We refer to them as $\dot{E}_{g}$ and $\dot{E}_{\phi}$ respectively. The luminosity depends on the scalar charge through the quantity
\be
\label{eq:alpha0}
\alpha_a^0 = \frac{d \ln m_a(\phi)}{d\phi} \,,
\ee
where the superscript ``0'' refers to the quantity being evaluated at $\phi_\infty$ (a constant corresponding to the scalar field at spatial infinity),
and where $m_a(\phi)$ is the effective $4d$ mass of a source $a$, which may depend on the dilaton.
In general, for a circular binary, the leading order term in  $\dot{E}_{\phi}$ is dipolar and depends on the difference in scalar charges of the binary constituents~\cite{Damour:1992we}. 

In our compactified (Kaluza-Klein) higher-dimensional gravity picture, the effective $4d$ mass of source $a$ is given by
\be
m_a(\phi)=m_a e^{-\phi/\sqrt{3}}\,,
\ee
as in \eqref{eq:m_eff}, and where we used the dilaton rescaling as in \eqref{eq:rescale}.
Since in our theory masses are coupled to the dilaton universally, 
\be
\alpha_1^0=\alpha_2^0=-1/\sqrt{3}\, ,\label{ouralphas}
\ee
the dipole radiation is zero (because $\alpha_1^0-\alpha_2^0=0),$ and so the leading contribution in $\dot{E}_{\phi}$ is quadrupolar. Therefore, both $\dot E_g$ and $\dot E_{\phi}$ are quadrupolar, and so is their sum, which is in agreement with our earlier findings \eqref{eq:luminosity}. More precisely, given the  Kaluza-Klein scalar charges \eqref{ouralphas},  the leading order contribution to the luminosity in Einstein-Maxwell-dilaton theories ~\cite{Julie:2018lfp,Khalil:2018aaj} becomes
\ba
\dot E_g \simeq  \left( 1 + \alpha_1^0  \alpha_2^0  \right)^{{-1}}\dot{E}_\GW^{(4)}\,, \qquad \dot E_\phi \simeq   \frac{1}{6} \left( 1 + \alpha_1^0  \alpha_2^0  \right)^{{-1}} \alpha_1^0  \alpha_2^0\,\dot{E}_\GW^{(4)}\,. 
\ea
Thus, in total, we have
\ba
\dot E_g+\dot E_\phi &\simeq &  \left( 1 + \alpha_1^0  \alpha_2^0  \right)^{{-1} }\left( 1 +  \frac{1}{6}  \alpha_1^0  \alpha_2^0 \right)\dot{E}_\GW^{(4)} \nonumber \\
& = & {\frac{19}{24}} \dot{E}_\GW^{(4)}\,,
\ea
which matches with our result in \eqref{eq:luminosity}.

\section{Constraints From gravitational wave observations}\label{constraintsgw}


In this section we compute the phase of the gravitational waveform in the frequency domain and compare it with  observations. We restrict ourselves to the leading post-Newtonian contribution. 
We begin by deriving the frequency evolution of the gravitational waves from the energy-balance law
\be\label{eq:balance-law}
\frac{dE}{dt}=\dot{E}_\GW\,,
\ee
which simply states that the rate of change of the binding energy of the binary $E$ is same as the luminosity $\dot{E}_\GW$ of the energy radiated by gravitational waves. For a circular binary, the binding energy is same as the effective potential, which is given in \eqref{bin2}. However, since we are interested in calculating the leading post-Newtonian effect, we can ignore the exponentially suppressed correction. We can further use the Kepler's law to rewrite the binding energy as
\be\label{eq:binding-energy}
E=-\frac{1}{2}\mu (G_N M \Omega)^{2/3}\,.
\ee
Substituting \eqref{eq:luminosity} on the right hand side of \eqref{eq:balance-law}, and \eqref{eq:binding-energy} into its left hand side, we find
\be\label{eq:fdot}
\dot{f}^{(5,c)}=\frac{76}{5} \pi ^{8/3} f^{11/3} G_N^{5/3}\mathcal M^{5/3}\,,
\ee 
where $f=\Omega/\pi$ is the gravitational waves frequency (this is manifest in  \eqref{eq:binarypert1}), 
and $\mathcal{M}=(m_1m_2)^{3/5}/(m_1+m_2)^{1/5}$ denotes the chirp mass. On the other hand, the frequency evolution in $4d$ GR is given by
\begin{equation}
\dot f^{(4)} = \frac{96}{5} \pi ^{8/3} f^{11/3} G_N^{5/3} \mathcal M^{2/3}\,,
\end{equation}
which differs from the compactified $5d$ result in \eqref{eq:fdot} by a numerical factor independent of the size of the extra dimension.

We now compute the gravitational wave phase in the frequency domain. The observed waveform is given by a linear combination of the + and $\times$ modes. 
In stationary phase approximation, the phase of gravitational waveform as a function of the frequency $f$ is~\cite{Maggiore:1900zz,Cutler:1994ys}
\begin{equation}\label{eq:phase-fourier}
\Psi(f)=2\pi f \,t(f)-\varphi(f)-\frac{\pi}{4}\,,
\end{equation}
where
\ba\label{eq:tf}
t(f)=t_0+\int_\infty^f df\, \frac{dt}{df}  =t_0+\int_\infty^f df\, \frac{1}{\dot{f}}\,,
\ea 
and 
\ba\label{eq:phase}
\varphi(f)=\int dt\, 2 \pi f =\varphi_0+\int_\infty^f df\,\frac{2\pi f}{\dot{f}}\,.
\ea
Further using \eqref{eq:fdot}  we obtain
\be\label{eq:phase-extradim}
\Psi^{(5, c)}(f)=\frac{9}{304 G_N^{5/3} u^5}+2\pi f t_0-\varphi_0-\frac{\pi}{4},
\ee
where $t_0$ and $\varphi_0$ are the time and phase at the coalescence respectively, and $u\equiv(\pi \mathcal{M} f)^{1/3}$ is the effective relative velocity of the binary components. On the other hand, the $4d$ GR result for the phase of the gravitational waves in the frequency domain is~\cite{Maggiore:1900zz,Cutler:1994ys}
\be\label{eq:phase-GR}
\Psi^{(4)}(f)=\frac{3}{128 G_N^{5/3} u^5}+2\pi f t_0-\varphi_0-\frac{\pi}{4}\,.
\ee
Thus, we can rewrite \eqref{eq:phase-extradim} based on \eqref{eq:phase-GR} as
\be
\Psi^{(5,c)}(f)=\frac{3}{128 G_N^{5/3} u^5}(1 + \delta \hat{\varphi})+2\pi f t_0-\varphi_0-\frac{\pi}{4}\,,
\ee
with 
\begin{equation}
\label{eq:varphi0}
\delta\hat{\varphi} \equiv \frac{5}{19} \sim 0.26\,.
\end{equation}
We note that our results agree with those derived in the context of Einstein-Maxwell-dilaton theory discussed in Ref.~\cite{Khalil:2018aaj} (with $\alpha_1=\alpha_2=-1/\sqrt{3}$ and the electric charges set to zero as discussed previously) to $0.3\%$. The difference arises due to a series expansion in luminosity in Ref.~\cite{Khalil:2018aaj} which assumes that the scalar energy flux is small compared to the tensor energy flux. If one performs the calculation without making such an approximation, the result in Ref.~\cite{Khalil:2018aaj} matches with ours exactly.

Let us now compare the predictions of our model, with a compactified fifth dimension with actual gravitational wave observations.
From \eqref{eq:varphi0}, one sees that the leading post-Newtonian term in \eqref{eq:phase-extradim} differs by $26\%$ from that of \eqref{eq:phase-GR}, irrespective of the masses of the binary components. The LIGO/Virgo Collaborations used the events detected from the first  and second observing runs and have placed upper bounds on $|\delta\hat{\varphi}|$ as $\sim 15\%$ from single events and $\sim 10\%$ from combined events~\cite{LIGOScientific:2019fpa}. Hence a discrepancy of $26\%$ is inconsistent with the gravitational wave observations, and  thus we can rule out the simple compactified $5d$ GR model considered in this paper.

\begin{figure}[htb]
\includegraphics[width=9.5cm]{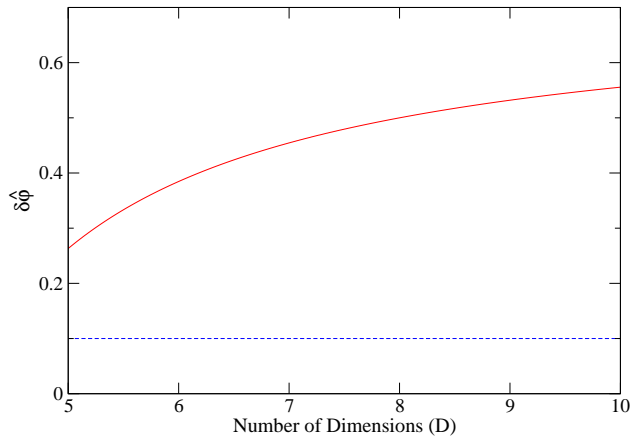}
\caption{The fractional difference ($\delta\hat{\varphi}$) of the GW phase with respect to that of the $4d$ GR as a function of the number of dimensions ($D$). The blue dashed line shows the upper bound placed on $\delta\hat{\varphi}$ from the combined events of the first and second observing runs of the  LIGO/Virgo~\cite{LIGOScientific:2019fpa}.
\label{fig:phaseD}}
\end{figure}

We can easily generalize our previous results and compute the phase of the gravitational waves in an arbitrary number dimensions $D$, with four non-compact dimensions and the rest compactified (periodic). Using the the gravitational wave luminosity in a $D$ dimensional spacetime given in footnote~\ref{EdotD}, it is straightforward to derive $\delta\hat \varphi$ as
\be
\delta\hat{\varphi}^{(D)}=\frac{5(D-4)}{7D-16}\,.
\ee
We plot $\delta\hat{\varphi}^{(D)}$ as a function of $D$ in Figure~\ref{fig:phaseD}, and notice that $\delta\hat{\varphi}$ increases with $D$. This means that our model stays inconsistent with the LIGO/Virgo observations even if we increase the number of compact extra dimensions.

We now comment on some caveat in the above bounds. We used the bounds derived by the LIGO/Virgo Collaborations that assumed the correction to 4d GR in the phase enters only at 0PN order. Such a correction partially degenerates with the chirp mass that also enters first at 0PN order, though the mass also enters at higher PN orders and thus the degeneracy can be partially broken. In reality, higher PN corrections to 4d GR also enter in the waveform phase. This may change the amount of correlation between the chirp mass and beyond-4d-GR effects and may weaken the bound on the 0PN correction. In~\cite{TheLIGOScientific:2016src}, the LIGO/Virgo Collaborations carried out another analysis for GW150914 where they included phase corrections at various PN orders in the search parameter set. This enhances the correlation significantly and the bound on the 0PN correction now becomes $|\delta\hat \varphi| \lesssim 5$. If we quote this bound, we cannot rule out the compact extra dimension model considered here. Thus, to make a robust statement on whether one can rule out the model with gravitational-wave observations, one needs to compute corrections at higher PN orders and rederive bounds on the extra dimension effect.

Having said this, one can still rule out the model with binary pulsar observations for the following reason. A standard method for testing GR with binary pulsars is to determine the masses from at least three independent observables (such as post-Keplerian parameters including the periastron precession, Shapiro delay and orbital decay rate) assuming GR and check the consistency. The orbital decay rate $\dot P$ is the only post-Keplerian parameter that depends on the gravitational-wave emission. Thus, even for the compact extra dimension model considered here, one can safely use the masses obtained from other post-Keplerian parameters under the 4d GR assumption  since the conservative corrections are exponentially suppressed. One can then use the measurement of $\dot P$ to constrain the model without having to worry about the degeneracy between the extra dimension effect and masses. Such $\dot P$ measurements have been mapped to a bound on $\delta\hat \varphi$ as $|\delta\hat \varphi| \lesssim 10^{-3}$~\cite{Yunes:2010qb,Nair:2020ggs}, which is much stronger than the gravitational-wave bound. Thus, one can rule out the compact dimension model with the binary pulsar observations\footnote{A similar result was found in~\cite{Durrer:2003rg} though this reference effectively introduces matter after the KK reduction and thus is different from the setup we study here.}.

\section{Conclusions}\label{Conclusions}

In this paper we performed an analysis of gravitational waves sourced by a binary in a $D$-dimensional spacetime with four non-compact dimensions and a set of compactified extra dimensions. We worked under the assumptions that the two binary sources are point-like and located on the same "brane" (i.e. at the same position in the compact coordinates). For most part we took $D=5$, but we have provided generalizations to arbitrary $D$ throughout the paper.
We worked within the framework of  GR, and in the limit of small extra dimensions. We computed the gravitational waves sourced by the binary, the luminosity of the gravitational waves and the phase of the gravitational waves, to leading order in the post-Newtonian expansion. We found that the luminosity of gravitational waves emitted in $5d$ gravity by a binary with the same characteristics (same masses and separation distance) as a $4d$ binary is 20.8\% less relative to the $4d$ case, to leading post-Newtonian order. The phase of the gravitational waveform  differs by 26\% relative to the $4d$ case, to leading post-Newtonian order, while for general $D$, the fractional difference of the phase with that of $4d$ GR is $\frac{5(D-4)}{7D-16}$, which only increases with an increase in $D$. While there are exponential corrections which depend on the size of the extra dimensions, the leading order estimates for the gravitational wave phase we gave here are independent of size, and depend only on the number of extra dimensions.
Based on a comparison with gravitational-wave observations from the  LIGO/Virgo Collaborations \cite{LIGOScientific:2019fpa} and binary pulsar observations from radio astronomy~\cite{Yunes:2010qb,Nair:2020ggs} we can rule out this class of  models for compact extra dimensions. The main source of discrepancy is the higher-dimensional gravity coupling with matter, which, when seen from a $4d$ perspective, means that matter will couple not only with the $4d$ metric, but with the dilaton as well. This dilaton coupling (or scalar charge) is responsible for fifth force effects which change the phase of the gravitational waves. Our results agree with those derived in the context of $4d$ Einstein-Maxwell-dilaton theory~\cite{Julie:2017rpw,Julie:2018lfp,Khalil:2018aaj} provided that we set the  binary's scalar charges equal to one another and equal to the Kaluza-Klein value. 

The same fifth force effects are responsible for the difference between the $4d$ Newton's constant $G_{\N}$ and the $4d$ gravitational coupling $G$: $G_N=\frac{2(D-3)}{D-2}G$, and for the Shapiro time-delay discrepancy with $4d$ GR.
In a parametrized post-Newtonian expansion (PPN) the $4d$ ``physical" metric (which is obtained by performing a rescaling of the $4d$ metric with the dilaton in such a way to eliminate the matter-dilaton coupling \cite{will-living} is written as $g_{00}=-1+2 U+\dots, g_{ij}=\delta_{ij}(1+2\gamma\, U+\dots)$, with $\gamma=1$ in $4d$ GR. A measurement of the frequency shift of radio photons to and from the Cassini spacecraft as they passed near the Sun gave $ \gamma = 1 + (2.1 \pm 2.3) \times 10^{-5}$ \cite{Bertotti:2003rm}. On the other hand, the ``physical metric" as read off from footnote 1 has $g_{00}=-1+\frac 23 \tilde h_{\T}{}_{00} $ and 
$g_{ij}=1+\frac 13 \tilde h_{\T}{}_{00}+\dots$, which amounts to $\gamma \sim 1/2$. Therefore this class of compactified extra dimensions models was ruled out based on Solar System measurements~\cite{Xu:2007dc,Deng:2015sua}. 

In string theory the massless dilaton is one of the many moduli (zero mass scalars) that arise in  the compactification of the higher-dimensional spacetime. Stabilzation of the moduli can be achieved, for example, by turning on fluxes for the Ramond-Ramond potentials \cite{Kachru:2003aw}. This 
gives rise to a mass term for the moduli, and eliminates the large contribution of the scalar fifth force, by turning a Coulomb potential into a Yukawa potential. It would be interesting to study gravitational waves in such a set-up and place constraints on the various parameters.
A somewhat simpler scenario is the Randall-Sundrum model, where the fifth dimension is large and warped. Our work here is intended as a first step in understanding how to set up the problem of solving for gravitational waves in a higher-dimensional space with compact dimensions, or warped, large extra dimensions. For example, we saw that quadrupole formula need not apply, and we had to use a direct integration of the Einstein equations.  Also, when it comes to the propagation of gravitational waves in spacetimes with warped, large extra dimensions, reducing the problem to $4d$ seems less appropriate, and working in the higher-dimensional space, as we did here, presents an advantage.

\acknowledgments
We thank Eric Poisson and Leo Stein for useful comments.
The work of  D.V. and Y.D. was supported in part by  the U.S. Department of Energy under Grant No. DE-SC0007984. 
S.T. and K.Y. acknowledge support from the Ed Owens Fund. 
K.Y. would like to also acknowledge support from NSF Award PHY-1806776, a Sloan Foundation Research Fellowship, the COST Action GWverse CA16104 and JSPS KAKENHI Grants No. JP17H06358. 

\appendix

\section{Modified quadrupole formula}
\label{appendix: quadrupole formula}
Here we explicitly point out why the usual quadrupole formula 
\be
\int d^3\vec x \, \tau^{i j}  = \frac{1}{2} \partial_0^2 \int  d^3\vec x\,\tau^{00} x^i x^j\,  ,
\ee
cannot be directly extended to higher dimensions if there are compact dimensions.

Consider the expression $\frac{1}{2}\partial_0^2 \int d^3\vec x \int_0^l dw\,\tau^{00} x^I x^J $, then use repeatedly the conservation law for $\tau^{MN}$ and integrate by parts, while paying attention to the boundary terms: 

\ba
\frac{1}{2}\partial_0^2 \int d^3\vec x 
\int_0^l dw\, \tau^{00} x^I x^J
    =\int d^3\vec x \int_0^l dw\, \tau^{IJ}+\frac 12 \int d^3\vec x(\partial_K\tau^{5K} x^Ix^J - \tau^{5I} x^J -\tau^{5J} x^I )\Big|_{w=0}^{w=l}\,. \nonumber \\
\ea
 For $I,J=i,j=1,2,3$, the quadrupole formula applies because of the periodicity of the metric fluctuations in $w$. However if either $I$ or $J$ are along the compact dimension, then there are extra terms relative to those expected based on the quadrupole formula. These terms are not straightforward to evaluate, and for this reason we rely on direct integration of the Einstein equations. 
\section{Hidden Brane Scenario}\label{hidden}
We generalize  the Newtonian potential by having the observer located at $w=0$, and the static mass 
source $m$ on a hidden brane  at $w=w_1$:
\be\label{eq:A3}
T_{00}(x^\mu,w)=m\delta^3(\vec x)\delta(w-w_1)\,.
\ee
The corresponding Newtonian potential evaluated by the observer is
\begin{align}
\label{eq:V_Rw1}
 V^{(5,c)}(R,w_1)=-\frac{4}{3\pi} G^{(5)} m\sum_{n=-\infty}^{\infty} \frac{1}{R^{2}+\left(w_1+n l\right)^{2}}\,.
\end{align}
Using Poisson summation, the above sum can be evaluated to give the following result:
\begin{align}\label{eq:A4}
V^{(5,c)}(R,w_1)=&-\frac 43 \frac{G^{(5)}m}{l R}\tanh {\frac{2\pi R}{l}}\left(1+\operatorname{sech}\frac{2\pi R}{l} \cos \frac{2\pi w_1}{l}\right)\nonumber\\
&\simeq -\frac 43\frac{G^{(5)}m}{l R}\left(1+2 e^{-2\pi R/l} \cos \frac{2\pi w_1}{l}\right)\,,
\end{align}
where in the last step we used $R\gg l$ and $R\gg w_1$.
When $w_1=0$, \eqref{eq:A4} reproduces the Newtonian potential in Section \ref{sec:Kepler'sLaw} in the limit of $2\pi R\gg l$.

\section{Newtonian Potential from integrating the retarded  $5d$ compactified  Green's function}
\label{NpotintretG}

In this appendix we offer an alternative derivation of the Newtonian potential obtained in Section \ref{sec:Kepler'sLaw},
by using the retarded $5d$ compactified Green's function. A static source of mass $m$ located at the origin has
\be\label{eq:A1}
T_{00}(x^\mu,w)=m\delta^3(\vec x)\delta(w)\,.
\ee
Substituting ~\eqref{eq:L3c} and~\eqref{eq:A1} into ~\eqref{eq:L2} we obtain
\begin{align}\label{integral}
\tilde{h}_{\T,00}\left(x^\mu, w\right)&=-\frac{4}{\pi} G^{(5)} m \sum_{n=-\infty}^{\infty} \frac{1}{r_{n}} \int_{0}^{\infty}dt^{\prime}\, \frac{\partial}{\partial r_{n}}\frac{\theta\left(t-t^{\prime}-r_n\right)}{\sqrt{\left(t-t^{\prime}\right)^{2}-r_n^{2}}}\,,
\end{align}
where $r_n=\sqrt{x^2+y^2+z^2+(w-nl )^2}$. The integral in \eqref{integral} can be evaluated by re-expressing the $\partial_{r_n}$ derivative in terms of a time-derivative using
\begin{align}\label{eq:A2}
\frac{\partial}{\partial r_n} \left[\frac{\theta\left(\Delta t-r_{n}\right)}{\left(\Delta t^{2}-r_{n}^{2}\right)^{1 / 2}}\right]=&
-\frac{\partial}{\partial \Delta t} \left[\frac{\theta\left(\Delta t-r_{n}\right)}{\left(\Delta t^{2}-r_{n}^{2}\right)^{1 / 2}}\right]
+\frac{\theta\left(\Delta t-r_{n}\right)}{\left[(\Delta t)^{2}-r_{n}^{2}\right ]^{3 / 2}}\left(r_{n}-\Delta t\right)\,,
\end{align}
where $\Delta t=t-t^{\prime}$. Substituting \eqref{eq:A2} into \eqref{integral}  we obtain 
\begin{align}\label{eq:h00}
\tilde {h}_{\T}^{00}(x^\mu, w)=&-\frac{4}{\pi} G^{(5)} m \sum_{n=-\infty}^{\infty} \frac{1}{r_{n}} \int_{r_n}^{\infty}d\Delta t\frac{r_n-\Delta t}{\left[(\Delta t)^{2}-r_{n}^{2}\right ]^{3 / 2}}
=\frac{4}{\pi} G^{(5)} m \sum_{n=-\infty}^{\infty} \frac{1}{r_{n}^{2}}\,,
\end{align}
and the corresponding Newton's potential $V^{(5,c)}\equiv (-1/3) \tilde{h}_{\T,00}$ matches with the one in Eq.~\eqref{eq:M3}.



\section{Retarded Green's function}\label{Green}

The massless scalar $D$-dimensional flat space Euclidean Green's function $\mathcal G_{\rm E}(x, x')=\mathcal G_{\rm E}(x-x')$ is the inverse of the $D$-dimensional Laplacian
\be
\Delta_{(D)} \mathcal G^{(D)}_{\rm}E(x- x')=\delta^D(x-x'), \qquad \mathcal G^{(D)}_{\rm E}(x-x')=-\int \frac{d^D p}{(2\pi)^D}\frac{e^{i p\cdot x}}{p\cdot p}.
\ee
Starting from the unique Euclidean Green's function, in Minkowski signature the retarded Green's function is obtained via the analytical continuation $p^0_{\rm E}\longrightarrow -i(p^0+i\epsilon)$. The Euclidean metric $\delta^{MN}$ gets replaced by the Minkowski metric $\eta^{MN}$, and the $i \epsilon$ prescription yields the {\it retarded} Green's function:
\be
\mathcal G^{(D)}(x-x')=-\int \frac{d^D p}{(2\pi)^D}\,\frac{e^{i p\cdot x}}{-(p^0+i\epsilon)^2+p^i p^i}.
\ee
The location of the poles is in the lower half-plane, when viewed as a function of  $p^0$ as a complex variable.  To evaluate the integral one integrates over $p^0$, using Cauchy's theorem, and if $t-t'>0$, one picks up the contribution from the two poles, otherwise the retarded Green's function is zero.
If $D=4$, we recover the familiar  expression 
\be
\mathcal G^{(4)}(x-x')=-\theta(t-t')\int \frac{d^3 p}{(2\pi)^3}\frac{\sin( p (t-t'))} p 
e^{i \vec p\cdot (\vec x-\vec x')}=-
\theta(t-t')\frac{1}{4\pi r}\delta((t-t')-|\vec x-\vec x'|),
\ee
where we used $p$ to denote the magnitude of the spatial vector $\vec p$, i.e. $p=|\vec p|$.
If $D=5$, then
\begin{align}
\mathcal G^{(5)}(x-x')&=-\theta(t-t')\int \frac{d^4 p}{(2\pi)^4}\frac{\sin(p (t-t'))} p 
e^{i \vec p\cdot (\vec x-\vec x')}\nn\\&=-\theta(t-t')\frac{4\pi}{(2\pi)^4}\int_0^\infty dp \,p^3 \int_0^\pi d\theta\,\sin^2 \theta\,
\frac{\sin( p (t-t'))} p 
e^{i p|\vec x-\vec x'|\cos\theta}
\nonumber\\
&=-\theta(t-t')\frac{1}{4\pi^2|\vec x-\vec x'|}\int_0^\infty dp \, p \sin( p (t-t'))J_1(p|\vec x-\vec x'|)
\nn\\
&=-\theta(t-t')\frac{1}{4\pi^2  |\vec x-\vec x'|}\frac{\partial}{\partial |\vec x-\vec x'|} \int_0^\infty dp \, \sin( p (t-t'))J_0 (p|\vec x-\vec x'|)
\nn\\
&=-\theta(t-t')\frac{1}{4\pi^2 |\vec x-\vec x'|}
\frac{\partial}{\partial |\vec x-\vec x'|} \frac{\theta(t-t'-|\vec x-\vec x'|)}{\sqrt{(t-t')^2-(\vec x-\vec x')^2}}\,.
\end{align}

\section{Direct Integration vs. Quadrupole Formula}\label{App:DirectvsQuad}
Here we use post-Newtonian order counting to explain a somewhat subtle aspect of our calculations. For simplicity's sake, in this appendix we restrict ourselves to $4d$ GR. 
Specifically, we will show that if in computing the spatial metric fluctuations $\tilde h^{ij}$, one uses the quadrupole formula, then one can safely neglect nonlinear source terms in the relaxed Einstein equations. However, if one directly integrates the relaxed Einstein equations, then the nonlinear terms cannot be neglected already at the leading post-Newtonian order. Since our goal is only to highlight  the dependence on velocities, we will write our equations with squiggle lines, signaling that we are imprecise about numerical factors.

The relaxed Einstein equations in the harmonic gauge are given in \eqref{eq:L0}, whose $4d$ solution is given by
\be\label{eq:GWpert4d}
 \tilde h^{\mu\nu} \sim \frac{4}{R} \int  d^3\vec x \, \tau^{\mu\nu} (\vec x, t-R) \,,
\ee
where we assumed that we are working in the far-field approximation, $R$ is the distance from the source to the field point, and  the right hand side is  evaluated at the retarded time.

\subsection{Direct Integration}
\label{sec:DI}
We are interested in extracting the order of magnitude in a post-Newtonian (PN) expansion estimates of the spatial components $\tilde h^{ij}$ for a compact binary. We will be somewhat careless about the indices and numerical factors since we only care about counting the PN order, namely powers of the relative velocity of the binary constituents, $v$. Eq.~\eqref{eq:GWpert4d} should receive contributions from both $T^{ij}$ and $t_{\LL}^{ij}$. To leading PN order, the contribution from $T^{ij}$ is roughly given by
\be
\label{eq:h_T}
\tilde h^{ij}_{\T} \sim \frac{\mu}{R} v^i v^j \sim \frac{\mu}{R} v^2\,,
\ee
where $\mu$ is the reduced mass of the binary. To consider the contribution of $t_{\LL}^{ij}$, let us substitute $\tilde g^{ij}=\eta^{ij}-\tilde h^{ij}$ into \eqref{eq:landau1} and look at one term (inside $t_{\LL}^{ij}$), for example
\be
(-g) t_{\LL}^{ij} \sim \tilde h^{00}{}^{,i}\tilde h_{00}{}^{,j} \sim \tilde h_{00}{}_{,i}\tilde h_{00}{}_{,j}\,,
\ee
so that
\be
\label{eq:h_t}
\tilde h^{ij}_{\tl} \sim \frac{1}{R} \int  d^3x\, \tilde h_{00}{}_{,i}\tilde h_{00}{}_{,j} \,.
\ee

To compute \eqref{eq:h_t}, let us consider partitioning the spacetime into a near zone (NZ) and a far zone (FZ)\footnote{FZ is also called the radiation zone or the wave zone.} relative to the location of the sources. NZ is the region centered around the source with the size of the gravitational wavelength while FZ is the region exterior to it~\cite{Pati:2000vt}. Within the NZ, the gravitational fields can be considered as almost instantaneous and retardation can be neglected. The integral in \eqref{eq:h_t} can be decomposed into the NZ and FZ integrals. It turns out that the former dominates the latter, so what we need is the NZ solution for $\tilde h_{00}$. For a compact binary, the leading NZ solution is given by~\cite{Blanchet:1998vx}
\ba
h_{00}^\mathrm{NZ} &\sim & \frac{2m_1}{r_1} + \frac{2m_2}{r_2}\,, \\
 h_{ij}^\mathrm{NZ} &\sim & \left( \frac{2m_1}{r_1}  + \frac{2m_2}{r_2} \right) \delta_{ij}\,,
\ea
with $r_a \equiv |\vec x - \vec x_a|$ and where $a=1,2$ corresponding to one of the two sources. Thus, $h^\mathrm{NZ} \sim 4m_1/r_1 + 4 m_2/r_2$ and 
\be
\tilde h_{00}^\mathrm{NZ} \sim h_{00}^\mathrm{NZ} - \frac{1}{2} h^\mathrm{NZ} \eta_{00} \sim  \frac{4m_1}{r_1} + \frac{4m_2}{r_2}\,.
\ee

We now substitute the above equation into the right hand side of \eqref{eq:h_t}. Those terms that only depend on $r_1$ or $r_2$ will diverge and be dropped upon regularization, so what matters is the cross term between sources 1 and 2. Ignoring numerical factors, we find
\ba
\label{eq:hij-t}
\tilde h^{ij}_{\tl} &\sim& \frac{m_1m_2}{R} \int_\mathrm{NZ} d^3 x\, \partial_i \left( \frac{1}{r_1} \right) \partial_j \left( \frac{1}{r_2} \right)  + (1 \leftrightarrow 2) \nn \\
&\sim& \frac{m_1m_2}{R} \int_\mathrm{NZ} d^3 x\,\partial_i^{(1)} \left( \frac{1}{r_1} \right) \partial_j^{(2)} \left( \frac{1}{r_2} \right)  + (1 \leftrightarrow 2) \nn \\
&\sim& \frac{m_1m_2}{R} \partial_i^{(1)} \partial_j^{(2)} \int_\mathrm{NZ} d^3 x\,\frac{1}{r_1r_2}  + (1 \leftrightarrow 2)\,, 
\ea
where we changed the partial derivatives to source-derivatives $\partial_i^{(a)} \equiv \partial/\partial x^i_a$ so that derivatives can be taken outside of the integral. The remaining integral, is given by~\cite{Blanchet:2001aw} 
\be
\int_\mathrm{NZ}d^3 x\, \frac{1}{r_1r_2}  = -2\pi r_{12}\,,
\ee
where $r_{12}= |\vec x_1 - \vec x_2|$ is the binary separation. Thus,
\ba
\tilde h^{ij}_{\tl} &\sim& \frac{m_1m_2}{R} \partial_i^{(1)} \partial_j^{(2)}r_{12}  + (1 \leftrightarrow 2) \nn \\
&\sim & \frac{m_1m_2}{R} \frac{1}{r_{12}}\nn \\
&\sim & \frac{\mu}{R} \frac{M}{r_{12}}\nn \\
&\sim & \frac{\mu}{R} v^2\,,
\ea
where $M$ is the total mass and we used the Kepler's law in the last equation. This scaling could in fact be easily obtained from the first line of \eqref{eq:hij-t} by replacing all the length scales inside the integral with $r_{12}$. Notice that $\tilde h^{ij}_{\tl}$ is of the same PN order as $\tilde h^{ij}_{\T}$. This means that the contribution from $t_{\LL}^{ij}$ cannot be neglected even at the leading order.

\subsection{Quadrupole Formula}

So far we have seen that by using direct integration of the relaxed Einstein equations, where $\tilde h^{ij}$ is sourced by $\tau^{ij}$, both the linearized $\tilde h^{ij}_{\T}$ and second order in fluctuation $\tilde h^{ij}_{\tl}$ are of the same order in a velocity expansion.
However, if we replace our starting point for the derivation of  $\tilde h^{ij}$ with the quadrupole formula (see Appendix \ref{appendix: quadrupole formula})
\be
\label{eq:identity}
\int \tau^{ij} d^3x = \frac{1}{2} \frac{d^2}{dt^2} \int  d^3x\, \tau^{00} x^i x^j\,,
\ee
then we would be using $\tau^{00}$ in extracting $\tilde h^{ij}$. In this case, the contribution from $t_{\LL}^{00}$ is subleading to that of $T^{00}$, as we will now show.

Using \eqref{eq:identity} and \eqref{eq:GWpert4d} one derives
\be
\tilde h^{ij} \sim \frac{2}{R}  \frac{d^2}{dt^2} \int  d^3x\, \tau^{00} x^i x^j\,.
\ee
The contribution from $T^{00}$ has the same scaling as in \eqref{eq:h_T}, i.e.
\be
\tilde h^{ij}_{(T^{00})} \sim \frac{\mu}{R} v^2\,.
\ee
 On the other hand, the contribution from $t^{00}$ is given by
\ba
\tilde h^{ij}_{(t^{00})} &\sim & \frac{1}{R} \frac{d^2}{dt^2} \int_\mathrm{NZ} d^3x \,\tilde h_{00}^\mathrm{NZ}{}_{,i}\tilde h_{00}^\mathrm{NZ}{}_{,j} \, x^i x^j\nn \\
&\sim & \frac{1}{R} \frac{d^2}{dt^2} \int_\mathrm{NZ} d^3x \, \partial_i \left( \frac{m_1}{r_1} \right) \partial_j \left( \frac{m_2}{r_2} \right) \, x^i x^j \nn \\
& \sim & \frac{1}{R} \Omega^2 \frac{m_1}{r_{12}^2} \frac{m_2}{r_{12}^2} r_{12}^2 r_{12}^3 \nn \\
& \sim & \frac{m_1m_2}{R} r_{12} \Omega^2 \nn \\
& \sim & \frac{\mu}{R} \frac{M}{r_{12}} (r_{12} \Omega)^2  \nn \\
&\sim &  \frac{\mu}{R} v^4\,,
\ea
where $\Omega$ is the binary angular frequency and we replaced all the length scale by $r_{12}$  in the third line as noted in Sec. \ref{sec:DI}. Notice that $\tilde h^{ij}_{(\tl^{00})}$ is of higher order in velocities than  $\tilde h^{ij}_{(\T^{00})}$. Thus, once the expression for $\tilde h^{ij}$ is turned into the quadrupole formula, the contribution from $t^{00}$ becomes subdominant and one only needs to consider $T^{00}$ to leading order.
\section{GW luminosity in Transverse-Traceless (TT) Gauge}\label{app:TT}

The transverse traceless (TT) gauge for linearized gravitational fluctuations about a flat spacetime imposes the following conditions:\footnote{The TT gauge uses the residual gauge freedom of the Lorenz gauge to impose the additional conditions: $(\tilde h^{\TT})^M{}_M=0, \;\tilde h^{\TT}_{0M}=0$. Take $n_I$  to be pointing in the direction of propagation of the waves, assuming they are plane waves:
$\tilde h ^{\TT}_{MN}= \tilde h^{\TT}_{MN}(t- n_I x^I).$ Then, from  the harmonic gauge one finds that $n_I (\tilde h^{\TT})^{IJ}=\partial_I( \tilde h^{\TT})^{IJ}=0, (\tilde h^{\TT})^I{}_{I}=0.$ For spherical waves these relations remain true to leading order in $1/R$.}
\ba 
\tilde{h}^{\TT}_{0M}=0\,,\qquad (\tilde{h}^{\TT})\indices{^I_I}=0\,,\qquad \partial_J(\tilde{h}^{\TT})^{IJ}=0\,.
\ea
Starting from the trace-reversed metric fluctuations, $\tilde h_{IJ}$,  one can show that the transverse traceless components are obtained by simply acting with a transverse-tracelss projector
\ba 
\Lambda_{IJKL}=P_{IK}P_{JL}-\frac{1}{D-2}P_{IJ}P_{KL}\,,\label{biglambda}
\ea
and where $P_{IK}$ are projectors orthogonal to $n^I$, with $n^I$  the direction of propagation of the waves and $n^I n^I=1$:
\ba
P_{IJ}=\delta_{IJ}-n_In_J\, ,\qquad P_{IJ} n^J=0\, .
\ea
Therefore, the non-vanishing  (same as the trace-reversed) metric fluctuations in the TT gauge are
\ba 
\tilde h_{IJ}^{\TT}=\Lambda_{IJKL}\tilde h^{KL}\, ,\qquad (\tilde h^{\TT} )^I{}_I=0\, , \qquad n^I \tilde h_{IJ}^{\TT}= \tilde h_{IJ}^{\TT} n^J=0\,.
\ea

It is easy to verify that
\ba\label{eq:TT1}
\tilde h^{\TT}_{IJ}(\tilde h^{\TT})^{IJ}=&\tilde h_{IJ}\tilde h^{IJ}-2n^I n^J \delta^{KL} \tilde h_{IK}\tilde h_{JL}+\left(\dfrac{D-3}{D-2}\right) n^I n^J n^K n^L \tilde h_{IJ} \tilde h_{KL}\nn\\
&-\left(\dfrac{1}{D-2}\right) (\tilde h^{I}{}_I)^2+\left(\dfrac{2}{D-2}\right)n^K n^L \tilde h_{KL} \tilde h^I{}_I \, .
\ea
In particular, working in the TT gauge, the equation \eqref{eq:landau3} becomes simply
\be\label{eq:landau4}
t_{0K}n^K=-\frac{1}{32 \pi G^{(D)}}\left\langle \dot{\tilde h}^{\TT}_{IJ}(\dot{\tilde h}^{\TT})^{IJ}\right\rangle\,.
\ee
We can nevertheless recover \eqref{eq:landau5}, starting from the TT gauge expression \eqref{eq:landau4} and substituting \eqref{eq:TT1}, which is to be expected given that we are computing a gauge-invariant quantity.
\ew

\bibliography{ref}

\end{document}